\documentclass[showkeys,10pt,twocolumn,amsmath,showpacs,superscriptaddress,prb,aps]{revtex4-2}
\usepackage{graphicx}
\usepackage{bm}
\usepackage{xcolor}   
\usepackage{amssymb} 
\usepackage{float}
\usepackage{amsmath}
\usepackage[utf8]{inputenc}
\usepackage{hyperref}
\usepackage{siunitx}
\hypersetup{
    colorlinks=true,
    linkcolor=blue,
    citecolor=blue,
    filecolor=magenta,
    urlcolor=cyan,
}

\usepackage[normalem]{ulem}

\begin{document}

\title{Majorana signatures in an asymmetrically coupled quantum dot--topological superconducting nanowire junction}

\author{Levente M\'ath\'e}
\email[Corresponding author: ]{levente.mathe@itim-cj.ro}
\affiliation{National Institute for Research and Development of Isotopic and Molecular Technologies, 67-103 Donath, 400293 Cluj-Napoca, Romania}
\author{Ioan Grosu}
\affiliation{Faculty of Physics, Babeș-Bolyai University, 1 Kogălniceanu, 400084 Cluj-Napoca, Romania}
\author{Liviu P. Z\^arbo}
\affiliation{National Institute for Research and Development of Isotopic and Molecular Technologies, 67-103 Donath, 400293 Cluj-Napoca, Romania}
\author{Ionel \c{T}ifrea}
%\email[Corresponding author: ]{itifrea@fullerton.edu}
\affiliation{Department of Physics, California State University, Fullerton, CA 92831, USA}

\date{\today}

\begin{abstract}
We present a theoretical study of the quantum transport through a nanoscale system in which a central quantum dot (QD) is coupled asymmetrically to normal leads and to two Majorana bound states (MBSs) localized at the ends of a topological superconducting nanowire threaded by a tunable magnetic flux. The effects of the leads--QD coupling asymmetry parameter $\alpha$ and the bias voltage asymmetry parameter $q$ on the system's linear conductance are considered for the case of unhybridized and hybridized MBSs. In the zero-temperature limit, for unhybridized MBSs the system's linear conductance is finite only when the magnetic flux phase $\phi = (2n+1)\pi$ ($n\in\mathbb{Z}$) and it scales as $\mathcal{G}=2q\alpha e^2/[h(\alpha+1)]$, while for hybridized MBSs it presents a complicated dependence on the system's parameters. At finite temperature, for unhybridized MBSs, the system's linear conductance oscillates as a function of the magnetic flux phase $\phi$ with a period of $2\pi$, and the position of the linear conductance maxima can be shifted from $\phi=2n\pi$ to $\phi=(2n+1)\pi$ by simply varying the value of the bias voltage asymmetry parameter $q$. For hybridized MBSs, the conductance exhibits a similar behavior when the energy level of the central QD, $\varepsilon_d$, is tuned at the leads' Fermi level ($\varepsilon_d=\varepsilon_F$), although when $\varepsilon_d\neq\varepsilon_F$ the oscillation period changes to $4\pi$, and the position of the linear conductance maxima depends on the actual value of $\varepsilon_d$ and other parameters in the system. Our results highlight the experimental importance of the leads-QD and bias voltage asymmetry parameters, which are often present in realistic experimental setups, and can strongly affect the identification and observation of MBSs transport signatures.
\end{abstract}

%\keywords{quantum dot; Majorana bound states; charge transport}
%\pacs{}
\maketitle

\section{INTRODUCTION}
\label{sec:I}

One possible realization of fault-tolerant quantum computation in solid state systems involves Majorana bound states (MBSs) associated with quasiparticle excitations predicted to appear at the ends of a one-dimensional~\textit {p}-wave superconductor~\cite{Kitaev2001}. MBSs are exotic excitations similar to the Majorana fermions, particles that are their own antiparticles~\cite{Majorana1937}. In terms of second quantization, the creation and annihilation operators corresponding to MBSs are equal to each other. MBSs obey non-Abelian statistics, providing nonlocal electronic degrees of freedom fundamental for encoding robust quantum information~\cite{Kitaev2001,Kitaev2006}. This intrinsic nonlocality underpins their potential as a platform for topological quantum computation, offering robust protection against decoherence compared to conventional qubits~\cite{Kitaev2001,Kitaev2006,Nayak2008,Alicea2012,Leijnse2012,Beenakker2013,Aguado2017}. In addition to~\textit{p}-wave superconductors, MBSs were predicted to occur in the $\nu=5/2$ quantum Hall state~\cite{Moore1991}, in superconductors with exotic spin-triplet pairing symmetry~\cite{Volovik1999}, in topological insulators in proximity to a superconductor~\cite{Fu2008}, and more recently even in high-temperature superconductors~\cite{Mercado2022}.

An important characteristic of MBSs is that they appear at zero energy and are separated from other quasiparticle excitations in the system by a finite energy gap. Hence, direct measurements of electrical conductance in quantum dot (QD) systems coupled to a topological superconductor nanowire (Majorana nanowire) might be the path for the MBSs experimental identification~\cite{Flensberg2011,Liu2011}. For a QD symmetrically coupled to the normal leads and tuned to resonance, $\varepsilon_d=0$, Liu and Baranger showed that the zero-temperature, zero-bias conductance takes the characteristic value $e^2/(2h)$ in the topological superconducting phase, in contrast to $e^2/h$ in the trivial phase and zero conductance when the QD is coupled to a regular fermionic zero mode~\cite{Liu2011}. Unfortunately, one has to be careful when attributing peaks in electrical conduction solely to MBSs, as peaks in the conduction function can also arise from other zero-energy modes~\cite{Alicea2012}. For example, such contributions might come from Andreev bound states or disorder~\cite{Liu2017,Pan2020,Sarma2021,Valentini2022}. There have been substantial theoretical efforts that explored QD-Majorana nanowire setups to probe MBSs via various transport properties, including conductance~\cite{Cao2012,Lee2013,Cheng2014,Vernek2014,Stefanski2015,Ramos2018,Ueda2014,Dessotti2014,Jiang2015,Xia2015,Zeng2016a,Zeng2016b,Baranski2016,Schuray2017,Wang2018a,Wang2018b,Calle2020,Mathe2022,Mathe2023,Bahadou2025,Bahadou2026} and current noise~\cite{Cao2012,Lu2012,Chen2014a,Lu2014,Liu2015,Zhao2016,Lu2016}. 
Experimental efforts to detect these states have employed transport measurements in devices based on superconducting contacts and QD-Majorana nanowire junctions~\cite{Mourik2012,Finck2013,Deng2016,Sherman2016,Chen2017,Zhang2017,Guel2018,Deng2018,Lutchyn2018}.

Along with the regular electron tunneling between the QD nanostructure metal leads, the presence of the Majorana nanowire in the nanoscale device leads to additional physical processes, in particular to Andreev reflection~\cite{Law2009,Fu2010a,Wang2013a,Wang2013b}. This process might also interfere with the detection of MBSs in QD-Majorana nanowire systems. Different from the regular electron tunneling that involves the usual tunneling of an electron from one lead to the other lead via the intermediate QD, the local Andreev reflection process converts an electron from one lead into a hole in the same lead. An additional process, which involves the conversion of an electron from one lead to a hole in the other lead (crossed Andreev reflection), also occurs in this system~\cite{Zocher2013,Liu2013,Liu2014}. In general, both Andreev reflection processes influence the quantum charge transport in QD nanostructures coupled to MBSs~\cite{Klees2024,Sturm2025,Trocha2025}. QD--Majorana ring structures threaded by a magnetic field could provide a platform for studying Majorana physics, where quantum transport is tuned by flux-controlled operations and by adjusting the QD--MBSs coupling strengths~\cite{Liu2011,Prada2012,Zeng2016a,Zeng2016b,Ramos2018,Mathe2022,Yang2024,Bahadou2025,Yu2025,Bahadou2026}. Theoretical results predict that as a function of the magnetic flux, the linear conductance oscillates with a flux period of $h/e$ or $2h/e$ depending on the system's configuration~\cite{Mathe2022}. Experimentally, a $h/e$-periodic conductance has been observed in a Majorana island embedded in an Aharonov–Bohm interferometer~\cite{Whiticar2020}.

The quantum transport through a QD side-coupled to MBSs was previously considered in the presence of electron-phonon interaction~\cite{Mathe2022}. In the case of unhybridized MBSs, the zero-temperature linear conductance of the system was shown to exhibit a $2\pi$ periodicity as a function of magnetic flux phase, independent of the QD characteristic energy level or the QD-MBSs coupling strengths. In contrast, for a finite overlap between the MBSs, the periodicity of the linear conductance generally changed to $4\pi$ when the QD characteristic electron energy level was tuned away from the leads' Fermi level. In addition, when the MBSs hybridize, the differential conductance periodicity changes from $2\pi$ to $4\pi$ for finite electron-phonon coupling. These results suggest that the magnetic flux phase-induced periodic electron conductance can be used to provide insight into possible MBSs in the superconducting nanowire~\cite{Mathe2022}.

\begin{figure}[t]
	\includegraphics[width =1\linewidth]{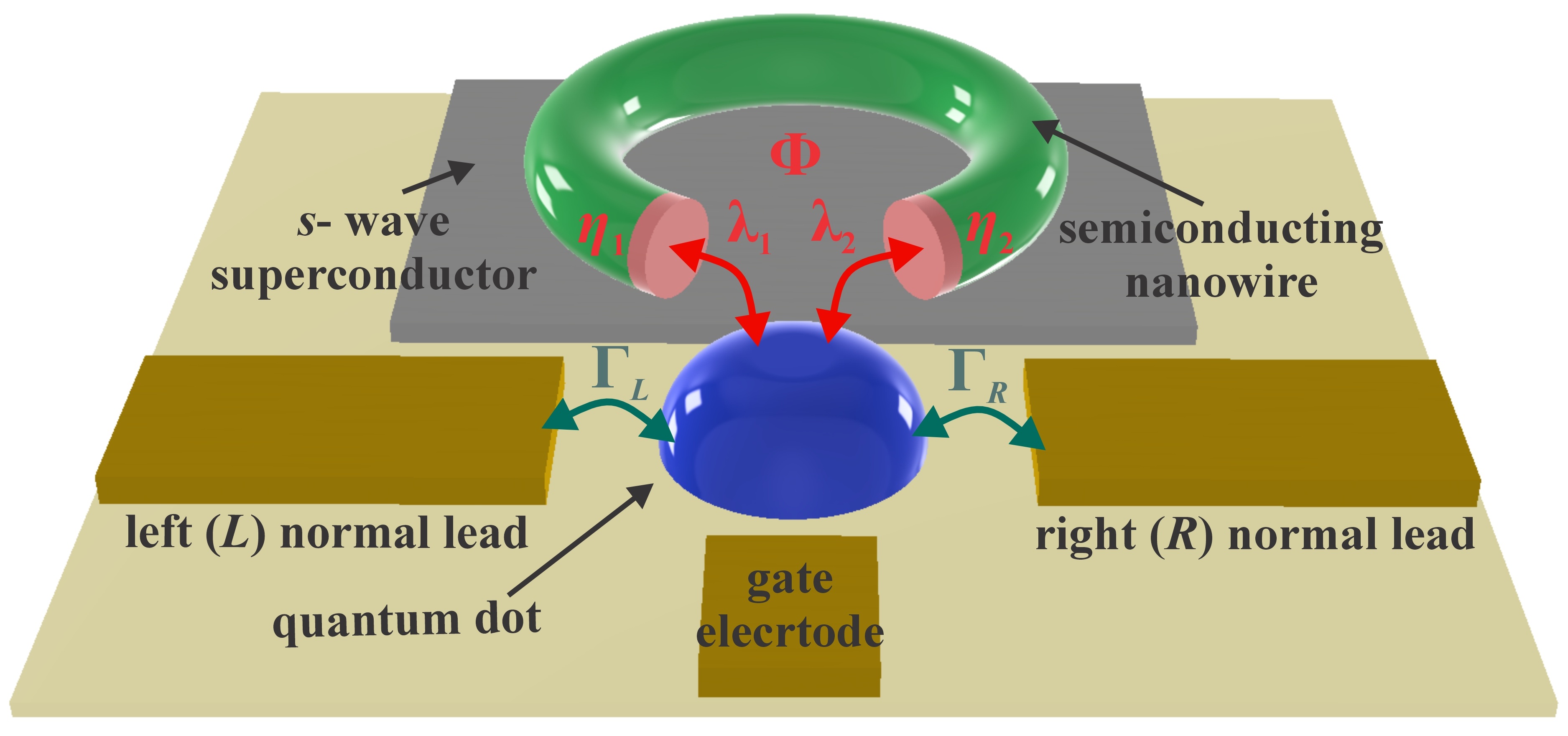}
	\centering
	\caption{Schematic representation of a QD attached to two MBSs hosted by a topological superconducting nanowire loop, threaded by a magnetic flux $\Phi$. The central QD is coupled to two metallic leads (left $L$ and right $R$) with coupling strengths $\Gamma_L$ and $\Gamma_R$, respectively. The central QD is also coupled to the two MBSs (denoted by the operators $\eta_1$ and $\eta_2$) located at the ends of a topological superconducting nanowire with coupling parameters $\lambda_1$ and $\lambda_2$, respectively. The central QD characteristic energy level can be tuned by a gate voltage applied to the gate electrode.}
	\label{fig:1}
\end{figure}

Here, we investigate the quantum transport in a QD system coupled to two MBSs induced in a topological superconducting nanowire loop threaded by a magnetic flux $\Phi$. The central QD is characterized by a single electron energy level $\varepsilon_d$ tunable by a gate voltage, and is connected to two normal metallic leads (left (L) and right (R)) that are subject to an applied external bias voltage (for a schematic representation of the system see Fig. \ref{fig:1}). We will consider the case in which the central QD is coupled to the two metallic leads asymmetrically, i.e., the two leads--QD coupling strengths are different, $\Gamma_L\neq\Gamma_R$. Additionally, the external bias voltage $\delta V$ is applied asymmetrically to the two leads. We will focus our analysis on the linear conductance of the system and examine how the leads--QD coupling and bias voltage asymmetries affect its properties. We will consider two scenarios of unhybridized and hybridized MBSs and identify possible signatures of these modes in the system's quantum transport, both in the zero and finite-temperature regimes. As a general result, we found that the system's characteristic parameters, such as the leads--QD and QD--MBSs coupling strengths, the MBSs overlap energy, and the QD characteristic energy level, play an important role in the system's charge quantum transport.

The article is organized as follows. In Sec.~\ref{sec:II}, we introduce our model and discuss the possible contributions to the system's charge quantum transport using the nonequilibrium Green's functions technique. Sec.~\ref{sec:III}  is dedicated to a numerical study of our results and a discussion of possible MBSs signatures in the system's linear conductance. Finally, Sec.~\ref{sec:IV} provides our conclusions and identifies possible future directions for quantum transport in QD systems coupled to MBSs.

\section{Theoretical model}
\label{sec:II}

We consider a system in which a central QD is laterally coupled to two MBSs located at the ends of a topological superconducting nanowire forming a loop threaded by a magnetic flux $\Phi$, with coupling parameters $\lambda_1$ and $\lambda_2$. The central QD can be accessed through two metallic electrodes, $L$ and $R$, kept at different chemical potentials, $\mu_L$ and $\mu_R$, and different temperatures, $T_L$ and $T_R$ (see Fig.~\ref{fig:1} for details). Theoretically, a semiconductor nanowire with strong Rashba spin--orbit coupling, subjected to a Zeeman field and proximity-coupled to a conventional $s$-wave superconductor, can host MBSs in the topological superconducting phase~\cite{Lutchyn2010,Oreg2010}. The magnetic flux $\Phi$ threading the loop controls the phase difference between the QD--MBSs coupling parameters. The topological phase is reached when the Zeeman energy satisfies the condition $E_Z>\sqrt{\Delta^2+\mu_w^2}$, where $E_Z=g\mu_B B/2$ ($g$ is the electron g-factor, $\mu_B$ is the Bohr magneton, $B$ is the magnitude of the applied magnetic field), $\Delta$ is the proximity-induced superconducting gap and $\mu_w$ is the chemical potential in the nanowire. In the low-energy transport regime considered here, the Zeeman energy is assumed to be much larger than the temperature, applied bias, lead--QD coupling strengths, and QD--MBSs coupling strengths. The system's Hamiltonian can be written as~\cite{Liu2011,Zeng2016a,Zeng2016b,Ramos2018,Mathe2022,Mathe2025a}:
\begin{equation}
    \mathcal{H}=\mathcal{H}_{\text{Leads}}+\mathcal{H}_{\text{MBS}}+\mathcal{H}_{\text{Dot}}+\mathcal{H}_{\text{Tun}}.
	\label{eq:1}
\end{equation}
Above, the first term in the Hamiltonian models the leads as noninteracting electron systems:
\begin{equation}
	\mathcal{H}_{\text{Leads}}=\sum_{\gamma,k}\varepsilon_{\gamma k}\,c_{\gamma k}^{\dagger}c_{\gamma k},
	\label{eq:2}
\end{equation}
where $c_{\gamma k}^{\dagger}$ and $c_{\gamma k}$ are the creation and annihilation operators for an electron with momentum \textit{k} in the lead $\gamma$ ($\gamma$ = $L$, $R$). The electron energy  $\varepsilon_{\gamma k} = \varepsilon_{k} - \mu_{\gamma}$ is measured relative to the lead's chemical potential $\mu_{\gamma}$. The second term in Eq.~\eqref{eq:1}, $\mathcal{H}_{\text{MBS}}$, describes the coupling between the two MBSs at the ends of the topological superconducting nanowire:
\begin{equation}
	\mathcal{H}_{\text{MBS}}=i \varepsilon_M \eta_1 \eta_2,
	\label{eq:3}
\end{equation}
with $\eta_1$ and $\eta_2$ the corresponding Majorana operators, and $\varepsilon_M \propto e^{-L/\xi}$ the Majorana overlap energy with $L$ and $\xi$ being the length of the Majorana nanowire and the superconducting coherence length, respectively. Due to the large Zeeman splitting, only one spin-polarized QD energy level is retained within the relevant transport-energy window. The QD is therefore modeled as a noninteracting single-level system:
\begin{equation}
	\mathcal{H}_\text{Dot}=\varepsilon_{d}d^{\dagger}d,
	\label{eq:4}
\end{equation}
where $\varepsilon_{d}$ is the QD characteristic energy level and $d^{\dagger}$ and $d$ are the creation and annihilation operators for an electron in the QD. Finally, the last term in the Hamiltonian describes the various tunneling processes in the system:
\begin{eqnarray}
	\mathcal{H}_\text{Tun}&=&\big[(\lambda_1 d - \lambda_1^* d^\dagger)\eta_1+ i (\lambda_2 d + \lambda_2^* d^\dagger)\eta_2\big]\notag\\
	&&+\sum_{\gamma,k}\big (V_{\gamma k}c_{\gamma k}^{\dagger}d +V^*_{\gamma k}d^{\dagger}c_{\gamma k}\big ).
	\label{eq:5}
\end{eqnarray}
The first two terms in Eq.~\eqref{eq:5} describe the coupling between the QD and the MBSs $\eta_1$ and $\eta_2$ located at the ends of the Majorana nanowire. The corresponding complex coupling parameters are $\lambda_1=|\lambda_1|e^{i\phi/4}$ and $\lambda_2=|\lambda_2|e^{-i\phi/4}$, respectively. The coupling strengths are given by $|\lambda_1|$ and $|\lambda_2|$, and $\phi=\pi \Phi/\Phi_0$ is the magnetic flux phase difference between the two coupling parameters ($\Phi$ is the magnetic flux from the threading magnetic field and $\Phi_0=h/(2e)$ is the magnetic flux quantum)~\cite{Zeng2016b,Mathe2022}. The last two terms in Eq.~\eqref{eq:5} model the coupling between lead $\gamma$ and the QD characterized by the $V_{\gamma k}$ tunneling amplitude. One can rewrite the Hamiltonian terms involving Majorana operators $\eta_1$ and $\eta_2$ by using their fermionic representation $\eta_1=(f^\dagger +f)/ \sqrt{2}$ and $\eta_2=i(f^\dagger -f)/ \sqrt{2}$, with  $f^\dagger$ and $f$ representing standard fermionic creation and annihilation operators.

When the system is subject to an external bias voltage $\delta V$, the electrical current in lead $\gamma$ can be evaluated in the subgap regime, $e|\delta V|<\Delta$, using the definition $I_\gamma=-e\left<dN_\gamma/dt\right>$ ($e$ is the electron charge and $N_\gamma$ is the particle number operator in lead $\gamma$) and the standard nonequilibrium Green's function techniques:~\cite{Michalek2013,Michalek2015,Huang2019,Huang2020,Mathe2022,Klees2024,Sturm2025}:
\begin{equation}
	\begin{split}
	I_\gamma= \frac{e}{h}\int &d\varepsilon \Big\{\mathcal{T}_{\gamma {\bar\gamma}}^{\text{ET}}(\varepsilon)\big[f_\gamma^{e}(\varepsilon)-f_{\bar\gamma}^{e}(\varepsilon)\big]\\
	&+\mathcal{T}_{\gamma {\gamma}}^{\text{LAR}}(\varepsilon)\big[f_\gamma^{e} (\varepsilon)-f_\gamma^{h} (\varepsilon)\big]\\
	&+ \mathcal{T}_{\gamma \bar\gamma}^{\text{CAR}}(\varepsilon)\big[f_{\gamma}^{e} (\varepsilon)-f_{\bar\gamma}^{h} (\varepsilon)\big]\Big\},
	\end{split}
	\label{eq:6}
\end{equation}
where $\mathcal{T}_{\gamma {\bar\gamma}}^{\text{ET}}(\varepsilon)$ is the electron transmission function associated with the electron tunneling between the two leads $\gamma$ and $\bar{\gamma}$ via the central QD,  $\mathcal{T}_{\gamma \gamma}^{\text{LAR}}(\varepsilon)$ is the local Andreev reflection transmission function associated with the conversion of an electron from the lead $\gamma$ into a hole in the same lead, and $\mathcal{T}_{\gamma {\bar\gamma}}^{\text{CAR}}(\varepsilon)$ is the crossed Andreev reflection transmission function associated with the process that converts an electron from lead $\gamma$ into a hole in lead $\bar{\gamma}$. These transmission functions satisfy the Fisher-Lee relations~\cite{Fisher1981}:
\begin{eqnarray}
\mathcal{T}_{\gamma {\bar\gamma}}^{\text{ET}}(\varepsilon)&=&\Gamma^e_\gamma \Gamma^e_{\bar\gamma}|G_{d11}^r(\varepsilon)|^2\nonumber\\
\mathcal{T}_{\gamma \gamma}^{\text{LAR}}(\varepsilon)&=&\Gamma^e_\gamma\Gamma^h_{\gamma}|G_{d12}^r(\varepsilon)|^2\label{eq:7}\\
\mathcal{T}_{\gamma {\bar\gamma}}^{\text{CAR}}(\varepsilon)&=&\Gamma^e_\gamma\Gamma^h_{\bar\gamma}|G_{d12}^r(\varepsilon)|^2\nonumber\;,
\end{eqnarray}
where  $\Gamma_{\gamma}^{e(h)} = 2\pi \sum_k |V_{\gamma k }|^2 \delta (\varepsilon \mp \varepsilon _{\gamma k})$ is the coupling strength between the central QD and the metallic lead $\gamma$ for electrons ($e$) and holes ($h$), respectively. In the following, we will consider the wide-band limit assuming electron-hole symmetry in the system, $\Gamma^e_\gamma=\Gamma^h_\gamma\equiv\Gamma_\gamma$. The two Green's functions, $G_{d11}^r(\varepsilon)=\langle \langle d|d^{\dagger}\rangle \rangle_\varepsilon^r$ and $G_{d12}^r(\varepsilon)=\langle \langle d|d\rangle \rangle_\varepsilon^r$, represent the electron-electron and electron-hole parts of the QD retarded Green's function matrix $\mathbf{G}_d^r(\varepsilon)$. Finally, $f_{\gamma}^{e}(\varepsilon)=[e^{(\varepsilon-\mu_\gamma)/(k_B T_\gamma)}+1]^{-1}$ and $f_{\gamma}^{h} (\varepsilon)=1-f_{\gamma}^{e}(-\varepsilon)$ denote the Fermi-Dirac distribution functions for electrons and holes, respectively. The electron-electron and electron-hole retarded Green's functions can be calculated using the equation of motion technique~\cite{Mathe2020,Mathe2022,Mathe2025a,Mathe2025b}:
\begin{subequations}
	\begin{eqnarray}
		G_{d11}^{r}(\varepsilon)
        &=&\Big\{ (\varepsilon + \varepsilon_d + i \Gamma) - \Big[ |\lambda_1|^2 + |\lambda_2|^2\label{eq:8a}\\
		&&+\frac{2\varepsilon_M}{\varepsilon}|\lambda_1\lambda_2|\cos\frac{\phi}{2} \Big]K(\varepsilon) \Big\} R(\varepsilon)^{-1},\notag\\
		G_{d12}^{r}(\varepsilon)&=&(|\lambda_2|^2 e^{i\phi/2}-|\lambda_1|^2 e^{-i\phi/2})K(\varepsilon) R(\varepsilon)^{-1},
		\label{eq:8b}\;\;\;
	\end{eqnarray}
\end{subequations}
where
\begin{eqnarray}
&&R(\varepsilon)=(\varepsilon -  \varepsilon_d + i  \Gamma)(\varepsilon +  \varepsilon_d + i  \Gamma) \notag\\
		&&- 2(\varepsilon + i  \Gamma) (|\lambda_1|^2 + |\lambda_2|^2)K(\varepsilon) 
		\label{eq:9}\\
		&&+ \frac{4 | \lambda_1\lambda_2|^2}{\varepsilon}K(\varepsilon) \cos^2\frac{\phi}{2} + \frac{4\varepsilon_M}{\varepsilon}\varepsilon_d |\lambda_1\lambda_2|K(\varepsilon) \cos\frac{\phi}{2},
		\notag\;\;    
\end{eqnarray}
with $K(\varepsilon)=\varepsilon/(\varepsilon^2 - \varepsilon_{M}^2)$ and $\Gamma=(\Gamma_L+\Gamma_R)/2$, respectively.

In the following, we investigate the quantum transport in the system and focus on the case of asymmetric lead--QD coupling strengths $\Gamma_L\neq\Gamma_R$. The two leads are considered at the same temperature $T_L=T_R\equiv T$, but subject to an external bias voltage $\delta V$ that results in a chemical potential difference between the two leads, $\mu_L - \mu_R = e\delta V$. The superconductor is assumed to be grounded, with its chemical potential fixed at $\mu_S=0$~\cite{Mathe2022,Sturm2025}.
The applied voltage is distributed asymmetrically between the two normal leads, such that
$\mu_L=\varepsilon_F+q\,e\delta V$ and
$\mu_R=\varepsilon_F+(q-1)e\delta V$,
with $0\leq q\leq1$.
The equilibrium Fermi energy is taken as the energy reference, $\varepsilon_F=\mu_S=0$~\cite{Mathe2022}. In the limit of small voltages, we can use a Taylor series expansion for the electron and hole Fermi-Dirac distribution functions to obtain the linear conductance:
\begin{equation}
	\mathcal{G}=\frac{e^2}{h} \int d\varepsilon\left[ -\frac{\partial f(\varepsilon)}{\partial \varepsilon} \right]\mathcal{T}(\varepsilon),
	\label{eq:10}
\end{equation}
where $f(\varepsilon) = [e^{\varepsilon/(k_B T)}+1]^{-1}$ is the equilibrium Fermi-Dirac function ($\varepsilon_F=0$) and  
\begin{equation}
	\begin{split}
		\mathcal{T}(\varepsilon)\! =\! \mathcal{T}^{\text{ET}}_{LR}(\varepsilon) + 2q\, \mathcal{T}^{\text{LAR}}_{LL}(\varepsilon)
		+ (2q - 1)\, \mathcal{T}^{\text{CAR}}_{LR}(\varepsilon)
        \label{eq:11}
    \end{split}
\end{equation}
is the transmission function due to contributions from electron tunneling, and local and crossed Andreev reflections. For the discussion of our results, it is convenient to split the system's linear conductance into its corresponding components
\begin{equation}
    \mathcal{G} = \mathcal{G}^{\text{ET}}+\mathcal{G}^{\text{LAR}}+\mathcal{G}^{\text{CAR}},
    \label{eq:12}
\end{equation}
with
\begin{equation}
    \mathcal{G}^{\text{ET}}=\frac{e^2}{h}\frac{\Gamma_L \Gamma_R}{4k_B T}\int d\varepsilon \frac{|G^r_{d11}(\varepsilon)|^2}{\cosh^2{\big(\frac{\varepsilon}{2k_B T}\big)}},
    \label{eq:13}
\end{equation}
the electron tunneling component,
\begin{equation}
    \mathcal{G}^{\text{LAR}}=\frac{e^2}{h}2q\,\frac{\Gamma_L\Gamma_L}{4k_BT}\int d\varepsilon \frac{|G^r_{d12}(\varepsilon)|^2}{\cosh^2{\big(\frac{\varepsilon}{2k_B T}\big)}},
    \label{eq:14}
\end{equation}
the local Andreev reflection component, and 
\begin{equation}
    \mathcal{G}^{\text{CAR}}=\frac{e^2}{h}(2q - 1)\frac{\Gamma_L\Gamma_R}{4k_B T}\int d\varepsilon\frac{|G^r_{d12}(\varepsilon)|^2}{\cosh^2{\big(\frac{\varepsilon}{2k_B T}\big)}},
    \label{eq:15}
\end{equation}
the crossed Andreev reflection component. Equations~\eqref{eq:12}-\eqref{eq:15} show that the properties of the system's linear conductance depend on the products $\Gamma_L\Gamma_R$ and $\Gamma_L\Gamma_L$, the moduli squared of the two Green's functions $|G^r_{d11}(\varepsilon)|^2$ and $|G^r_{d12}(\varepsilon)|^2$, and the temperature. Additionally, the local and crossed Andreev reflection components of the system's conductance depend on the bias voltage asymmetry factor, $q$. The two products, $\Gamma_L\Gamma_R$ and $\Gamma_L\Gamma_L$, will be discussed in terms of the leads--QD coupling asymmetry parameter $\alpha=\Gamma_L/\Gamma_R$. The moduli squared of the two Green's functions, $|G^r_{d11}(\varepsilon)|^2$ and $|G^r_{d12}(\varepsilon)|^2$, will be influenced by the system's configuration and according to Eqs.~\eqref{eq:8a} and~\eqref{eq:8b} they are a function of the central QD characteristic energy level $\varepsilon_d$, the QD--MBSs coupling strengths $|\lambda_1|$ and $|\lambda_2|$, the magnetic flux phase $\phi$, and the Majorana characteristic overlap energy $\varepsilon_M$. Our analysis will consider the case of unhybridized and hybridized MBSs corresponding to $\varepsilon_M=0$ and $\varepsilon_M\neq 0$, respectively. The QD--MBS coupling strengths will be considered in the weak ($|\lambda_1|/\Gamma<1$ and $|\lambda_2|/\Gamma<1$) and strong coupling ($|\lambda_1|/\Gamma>1$ and $|\lambda_2|/\Gamma>1$ ) regimes. The magnetic flux phase $\phi$ can be controlled using magnetic fields and induces a periodicity in the system's linear conductance.

\begin{figure*}[ht]
	\includegraphics[width =1\linewidth]{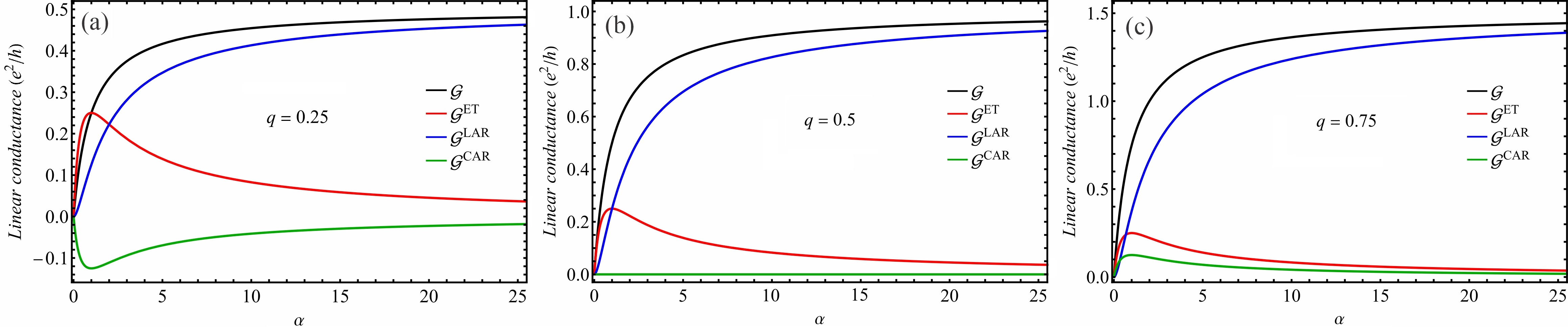}
	\centering
	\caption{The zero--temperature linear conductance $\mathcal{G}$ (black line) and its components related to electron tunneling $\mathcal{G}^{\text{ET}}$ (red line), local Andreev reflection $\mathcal{G}^{\text{LAR}}$ (blue line), and crossed Andreev reflection $\mathcal{G}^{\text{CAR}}$ (green line) as a function of the coupling asymmetry parameter $\alpha=\Gamma_L/\Gamma_R$ in the case of unhybridized MBSs ($\varepsilon_M/\Gamma=0$). The magnetic flux phase is set at $\phi=\pi$. The applied bias voltage asymmetry parameter is (a) $q=0.25$, (b) $q=0.5$, and (c) $q=0.75$.}
	\label{fig:2}
\end{figure*}

As a general result, in the linear transport regime, the electron tunneling contribution to the system's linear conductance does not depend on the bias voltage asymmetry parameter $q$, since it is driven by the electrochemical potential difference $\mu_L-\mu_R=e\delta V$. Differently, the Andreev reflection contributions to the linear conductance are both sensitive to the bias voltage asymmetry parameter $q$, i.e., $\mathcal{G}^{\mathrm{LAR}}\propto q$ and $\mathcal{G}^{\mathrm{CAR}}\propto (2q-1)$. In particular, this dependence leads to a sign change at $q=0.5$  for the crossed Andreev reflection contribution to the system's conductance ($\mathcal{G}^{\mathrm{CAR}}=0$ for $q=0.5$). An interesting situation arises in the case $0<q<0.5$, where the crossed Andreev reflection contribution to the system's linear conductance becomes negative. In particular, for $q\rightarrow 0$, the crossed Andreev reflection process becomes the dominating contribution in the total Andreev reflection process, a situation that can have implications for the experimental observation of this process in direct conductance measurements. On the other hand, when $0.5<q<1$, the local and crossed Andreev reflection contributions to the system's conductance are positive, making their identification in direct experimental measurement difficult.

The zero-temperature linear electron conductance can be evaluated using the Fermi-Dirac distribution function zero-temperature limit, $f(\varepsilon)=\theta(\varepsilon_F-\varepsilon)$, where $\theta(x)$ is the Heaviside step function. In this limit, the corresponding components of the system's linear conductance depend only on the values of the two Green's functions at $\varepsilon=\varepsilon_F=0$~\cite{Mathe2025a}:
\begin{equation}
    \mathcal{G}^{\text{ET}}=\frac{e^2}{h}\Gamma_L \Gamma_R |G^r_{d11}(0)|^2,
    \label{eq:16}
\end{equation}
\begin{equation}
    \mathcal{G}^{\text{LAR}}=2q\frac{e^2}{h}\Gamma_L\Gamma_L|G^r_{d12}(0)|^2,
    \label{eq:17}
\end{equation}
and
\begin{equation}
    \mathcal{G}^{\text{CAR}}=(2q-1)\frac{e^2}{h}\Gamma_L\Gamma_R|G^r_{d12}(0)|^2,
    \label{eq:18}
\end{equation}
respectively. The squared moduli of the Green's functions, $|G^r_{d11}(0)|^2$ and $|G^r_{d12}(0)|^2$, are periodic functions of the magnetic flux phase $\phi$, and this periodicity is directly reflected in the zero--temperature linear conductance. In the unhybridized case, $\varepsilon_M=0$, their characteristic period is $2\pi$, and they both present maxima at $(2n+1)\pi$ ($n\in\mathbb{Z}$), independent of the central QD characteristic energy level $\varepsilon_d$ and  QD--MBSs coupling strengths $|\lambda_1|$, and $|\lambda_2|$. 
In the hybridized case, $\varepsilon_M\neq 0$, $|G^r_{d12}(0)|^2$ vanishes regardless of the values of the central QD characteristic energy level $\varepsilon_d$, and the QD--MBSs coupling strengths $\lambda_1|$ and $|\lambda_2|$. On the other hand, for $\varepsilon_d=0$ and $\varepsilon_M\neq 0$, the period of the modulus squared of the Green's function $|G^r_{d11}(0)|^2$ is still $2\pi$, with maxima at the same points $(2n+1)\pi$ ($n\in\mathbb{Z}$), independent of the $|\lambda_1|$ and $|\lambda_2|$ values. Quite differently, for $\varepsilon_d\neq 0$, the periodicity of the $|G^r_{d11}(0)|^2$ function changes to $4\pi$, and the position of the local extreme points as a function of the magnetic flux phase $\phi$ will depend on the other parameters of the system \cite{Mathe2022}.

The evaluation of the system's linear conductance in the finite temperature regime involves the whole energy spectrum of the two Green's functions $|G^r_{d11}(\varepsilon)|^2$ and $|G^r_{d12}(\varepsilon)|^2$. The analysis of the energy dependence of the moduli squared of the two Green's functions reveals a complex behavior as a function of the system's parameters in both the unhybridized and hybridized cases.

\section{Results and discussion}
\label{sec:III}

In the following section, we present our results for the total linear conductance in the system. We will focus on the general case when the system presents asymmetry in the applied bias voltage ($q\neq 0.5$), asymmetry in the leads--QD coupling strengths ($\Gamma_L\neq\Gamma_R$), and different magnitudes $|\lambda_1|$ and $|\lambda_2|$ for the QD--MBSs coupling strengths. For our estimations, we will scale all energy-related quantities to $\Gamma=(\Gamma_R+\Gamma_L)/2$. Two different scenarios can be envisioned for the two leads--QD coupling strengths, $\Gamma_L$ and $\Gamma_R$. One possibility is that $\Gamma_L$ and $\Gamma_R$ change simultaneously as their total sum $\Gamma$ is kept constant ~\cite{Krawiec2007,Schmitt2011,Aligia2015,Daroca2018a,Daroca2018b}. Another possibility is that one of the leads--QD coupling strengths is kept constant while the other is varied~\cite{Krawiec2002,Daroca2018b}. In the following, we will consider the first option, with the sum of the coupling strengths, $\Gamma$, held constant, and analyze the system's total linear conductance as a function of the ratio $\alpha=\Gamma_L/\Gamma_R$.

The particular case with symmetric leads-QD coupling strengths $\Gamma_L=\Gamma_R$ and bias voltage $q=0.5$ was previously considered by M\'ath\'e \textit{et al.}~\cite{Mathe2022}. As already mentioned, in that case, the crossed Andreev reflection process does not contribute to the total linear conductance in the system. The total linear conductance in the case of unhybridized MBSs, $\varepsilon_M=0$, was found to showcase a 2$\pi$ periodicity as a function of magnetic flux phase $\phi$, independent of the QD characteristic energy $\varepsilon_d$, or the finite values of the QD-MBSs coupling strengths $|\lambda_1|$ and $|\lambda_2|$. The periodicity of the system's linear conductance changes to 4$\pi$ for the case of hybridized MBSs, $\varepsilon_M\neq 0$, when the energy level in the QD is tuned away from the Fermi level energy value $\varepsilon_F=0$. In the zero-temperature limit, $T=0$, for the unhybridized MBSs case, both the electron tunneling and local Andreev reflection contributions to the system's linear conductance are oscillating with a period of 2$\pi$, having maxima at $\phi=(2n+1)\pi$ ($n\in \mathbb{Z}$) with magnitude $\mathcal{G}^{\text{ET}}=\mathcal{G}^{\text{LAR}}=e^2/(4h)$. Implicitly, the total linear conductance is equal to $\mathcal{G}=e^2/(2h)$ ~\cite{Mathe2022}. This important zero-temperature property is a consequence of the analytical behavior of the two Green's functions $G_{d11}^r(\varepsilon)$ and $G_{d12}^r(\varepsilon)$ in the limit $\varepsilon\rightarrow 0$.

For our calculations, we will use parameter values similar to those considered in Ref. \cite{Mathe2022}. The largest energy scale in the system is the external magnetic field induced Zeeman energy $E_Z$, along with the superconducting gap in the Majorana nanowire $\Delta$. For the $\text{InSb}$ nanowires used in experiments, the proximity-induced superconducting gap is typically of the order of $\Delta \sim 250~\mu\mathrm{eV}$~\cite{Mourik2012}. The QD--MBSs coupling strengths $|\lambda_j|$ and the leads--QD coupling strengths $\Gamma_\gamma$ are usually in the range of a few $\mu\mathrm{eV}$ in experimental measurements~\cite{Cao2012,Deng2016,Deng2018}.

\subsection{Unhybridized MBSs, \texorpdfstring{$\varepsilon_M=0$}{EM0}}

\begin{figure*}[ht]
	\includegraphics[width =0.95\linewidth]{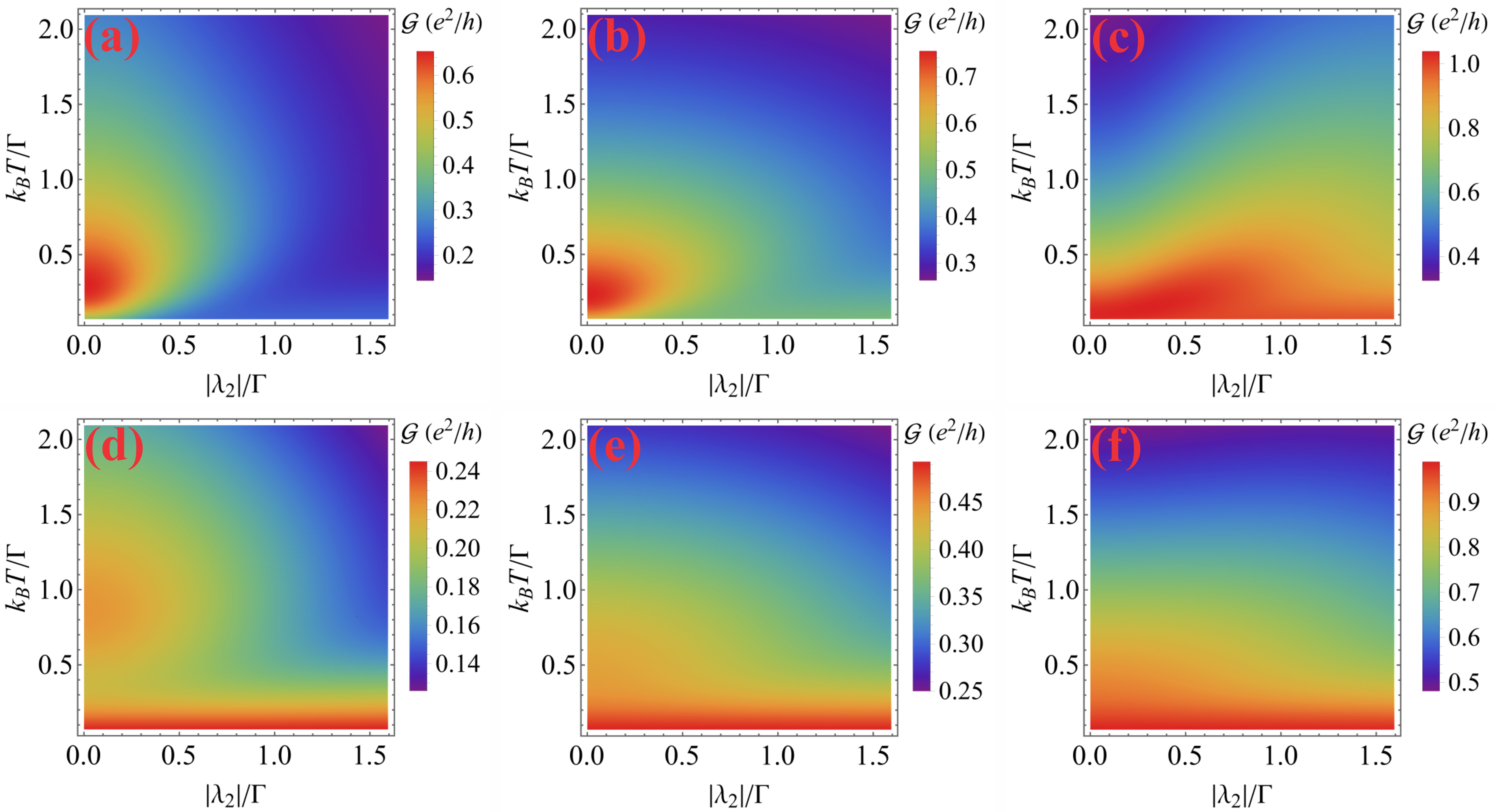}
	\centering
	\caption{The linear conductance $\mathcal{G}$ in the case of unhybridized MBSs ($\varepsilon_M/\Gamma= 0$) as a function of QD-MBS coupling strength $|\lambda_2|/\Gamma$ and temperature $k_B T/\Gamma$ for a symmetric leads--QD coupling parameter ($\alpha = 1$) at different values of the QD--MBS coupling strength $|\lambda_1|$ and bias voltage asymmetry parameter $q$. The QD characteristic energy level $\varepsilon_d/\Gamma = 0$, and the magnetic flux phase $\phi=\pi$. The values of the QD-MBS coupling strength $|\lambda_1|$ are: (a)-(c) $|\lambda_1|/\Gamma = 0.3$ and (d)-(f) $|\lambda_1|/\Gamma= 1.2$. The bias voltage asymmetry parameter is set to $q = 0.25$ for (a) and (d), $q=0.5$ for (b) and (e), and $q=1$ for (c) and (f).}
	\label{fig:3}
\end{figure*}

We will analyze first the system's linear conductance and its three components for the case of unhybridized MBSs, i.e., $\varepsilon_M=0$. In the zero-temperature regime, in the presence of an asymmetry in the leads-QD coupling strengths, $\alpha\neq 1$, one can easily evaluate the behavior of the two Green's functions $G^r_{d11}(\varepsilon)$ and $G^r_{d12}(\varepsilon)$ in the $\varepsilon\rightarrow 0$ limit. When the MBSs are unhybridized, the system's zero-temperature linear conductance vanishes as long as the magnetic flux phase in the Majorana nanowire is not an odd multiple of $\pi$, $\phi\neq (2n+1)\pi$ ($n\in\mathbb{Z}$). On the other hand, when the magnetic flux phase obeys the condition $\phi = (2n+1)\pi$, the electron tunneling, local and crossed Andreev reflection contributions to the system's linear conductance, at zero temperature, are given by 
\begin{eqnarray}
    \mathcal{G}^{\text{ET}}&&=\frac{e^2}{h}\frac{\alpha}{(\alpha + 1)^2},\label{eq:19}\\
    \mathcal{G}^{\text{LAR}}&&=2q\frac{e^2}{h}\frac{\alpha^2}{(\alpha + 1)^2},\label{eq:20}\\
    \mathcal{G}^{\text{CAR}}&&=(2q-1)\frac{e^2}{h}\frac{\alpha}{(\alpha + 1)^2},\label{eq:21}
\end{eqnarray}
respectively. Accordingly, the system's zero-temperature total linear conductance can be calculated as
\begin{equation}
    \mathcal{G} = 2q\frac{e^2}{h}\frac{\alpha}{\alpha + 1}\;. 
    \label{eq:22}
\end{equation}
Note that in this case, the system's linear conductance and its three different contributions are not dependent on the QD characteristic electron energy level, $\varepsilon_d$, and the QD-MBSs coupling strengths $|\lambda_1|$ and $|\lambda_2|$. As a function of the asymmetry parameter $\alpha$, all contributions to the system's conductance increase with $\alpha$, as long as $0<\alpha\leq 1$. When $\alpha>1$, the electron tunneling and the crossed Andreev reflection contributions to the system's linear conductance decrease as $\alpha$ increases. Differently, the local Andreev reflection contribution to the system's linear conductance increases as $\alpha$ increases. The system's setup symmetry implies that the electron tunneling and crossed Andreev reflection contributions to the system's linear conductance should be unaffected by a switch between the two external leads $L\leftrightarrow R$, a result confirmed by Eqs.~\eqref{eq:19} and~\eqref{eq:21}. As expected, in the case $\alpha \to 0$, the system's linear conductance $\mathcal{G} \to 0$, regardless of the applied bias voltage.

\begin{figure*}[ht]
	\includegraphics[width =0.85\linewidth]{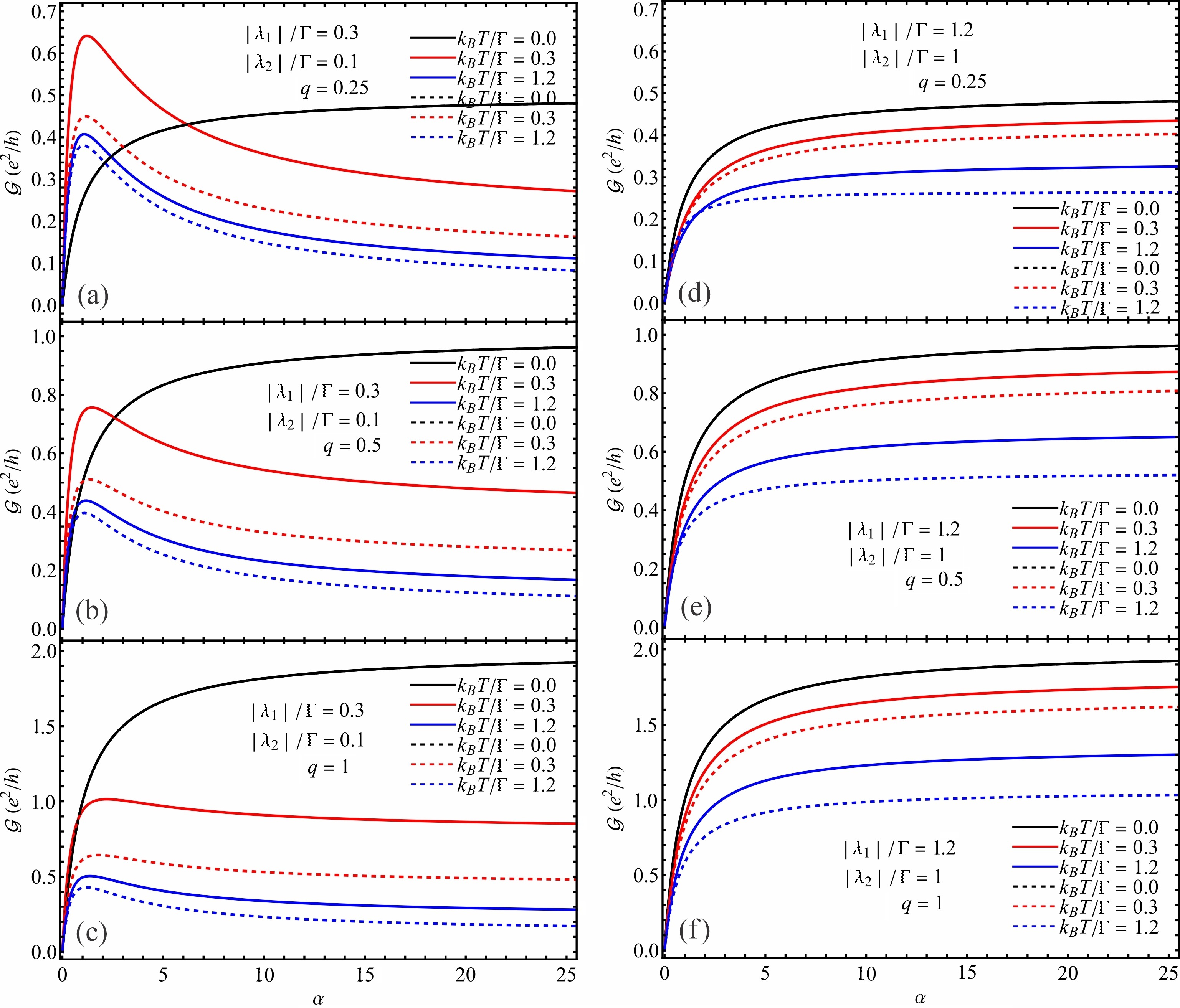}
	\centering
	\caption{Total linear conductance $\mathcal{G}$ for the case of unhybridized MBSs ($\varepsilon_M/\Gamma= 0$) as a function of the coupling asymmetry parameter $\alpha$ at different temperature values ($k_BT/\Gamma=0$ -- black line, $k_BT/\Gamma=0.3$ -- red line, and $k_BT/\Gamma=1.2$ -- blue line) and QD-MBS coupling strengths: (a)-(c) $|\lambda_1|/\Gamma = 0.3$, $|\lambda_2|/\Gamma=0.1$ and (d)-(f) $|\lambda_1|/\Gamma = 1.2$, $|\lambda_2|/\Gamma=1$. The bias voltage asymmetry parameter $q=0.25$ for (a) and (d), $q=0.5$ for (b), (e), and  $q=1$ for (c) and (f). The magnetic flux phase is $\phi=\pi$, and the QD characteristic energy level $\varepsilon_d/\Gamma = 0$ (solid lines) and $\varepsilon_d/\Gamma = -1$ (dashed lines).}
	\label{fig:4}
\end{figure*}

In the case of symmetric leads--QD coupling strengths, $\alpha=1$, the system's zero-temperature linear conductance components become $\mathcal{G}^{\text{ET}}=e^2/(4h)$, $\mathcal{G}^{\text{LAR}}=q e^2/(2h)$, and $\mathcal{G}^{\text{CAR}} = (2q-1) e^2/(4h)$, leading to a total linear conductance $\mathcal{G}=q\,e^2/h$. Under symmetric bias voltage, $q=0.5$, this reproduces the well-known result $\mathcal{G} = e^2/(2h)$, serving as a characteristic Majorana fingerprint~\cite{Liu2011}. In the limit $\alpha\rightarrow \infty$, the zero-temperature total linear conductance becomes $\mathcal{G}\simeq\mathcal{G}^{\text{LAR}}=2qe^2/h$. In this limit, the QD is effectively coupled to a single lead, and the system's linear conductance is dominated by local Andreev reflection. Note also that if we consider the $q\rightarrow 1$ limit along with $\alpha\rightarrow\infty$, the system's linear conductance becomes $\mathcal{G}\simeq\mathcal{G}^{\text{LAR}}=2e^2/h$ ~\cite{Wang2021}.

\begin{figure*}[t]
	\includegraphics[width =1\linewidth]{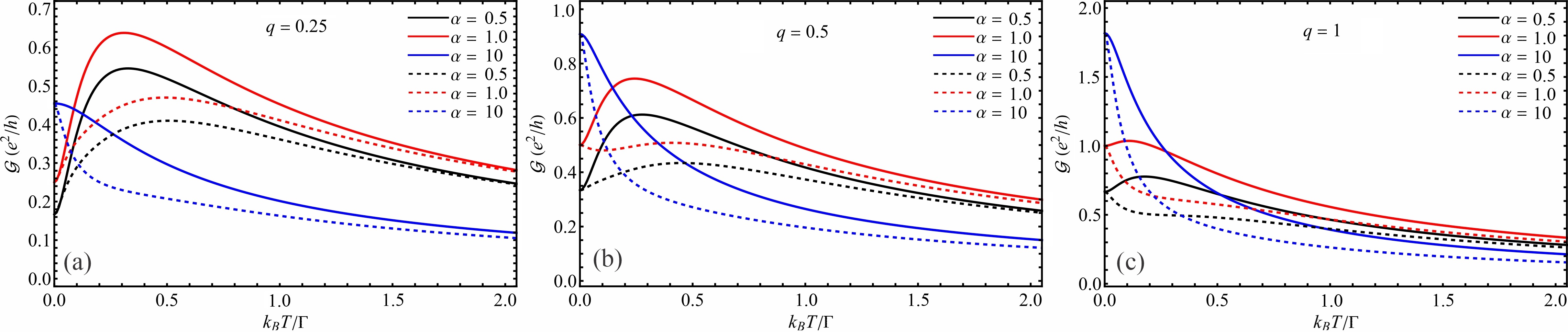}
	\centering
	\caption{The total linear conductance $\mathcal{G}$ in the case of unhybridized MBSs ($\varepsilon_M/\Gamma=0$) as a function of temperature $k_B T/\Gamma$ at different values of the leads--QD coupling asymmetry parameter ($\alpha=0.5$ -- black line, $\alpha=1$ -- red line, and $\alpha=10$ -- blue line) and different values of the bias voltage asymmetry parameter (a) $q=0.25$, (b) $q=0.5$ and (c) $q=1$. The QD-MBSs coupling strengths are $|\lambda_1|/\Gamma= 0.3$ and $|\lambda_2|/\Gamma = 0.1$, and the magnetic flux phase is $\phi=\pi$. The solid and dashed lines correspond to the QD characteristic energy level $\varepsilon_d /\Gamma = 0$ and $\varepsilon_d/\Gamma = -1$, respectively.}
	\label{fig:5}
\end{figure*}

Figures~\ref{fig:2}a--\ref{fig:2}c present the system's total linear conductance at zero temperature, $\mathcal{G}$ (black line), along with its electron tunneling, $\mathcal{G}^{\text{ET}}$ (red line), local Andreev reflection, $\mathcal{G}^{\text{LAR}}$ (blue line), and crossed Andreev reflection  $\mathcal{G}^{\text{CAR}}$ (green line) components as a function of the asymmetry parameter $\alpha$ for different values of the biasing parameter ($q=0.25$ for Fig.~\ref{fig:2}a, $q=0.5$ for Fig.~\ref{fig:2}b, and $q=0.75$ for Fig.~\ref{fig:2}c). The MBSs in the system are considered to be unhybridized, $\varepsilon_M=0$, and the magnetic flux phase is $\phi=\pi$. As we already discussed, other parameters in the model are irrelevant in the zero--temperature regime for the unhybridized case. The electron tunneling and the crossed Andreev reflection contributions to the system's conductance vanish as $\alpha$ increases beyond $\alpha=1$, independent of the applied bias voltage value. Consequently, in the large $\alpha$ limit, the system's total linear conductance is mainly due to the local Andreev reflection process. The electron tunneling contribution to the system's conductance presents a maximum at $\alpha=1$ independent of the applied bias voltage asymmetry, $q$. The crossed Andreev reflection contribution to the system's linear conductance changes sign independently of the asymmetry parameter $\alpha$ when the asymmetry parameter in the applied bias voltage, $q$, crosses the $q=0.5$ value. Additionally, the crossed Andreev reflection contribution presents a minimum at $\alpha=1$ for $0<q<0.5$, and a maximum at $\alpha=1$ for $0.5<q<1$. Also, $\mathcal{G}^{\text{CAR}}=0$ for $q=0.5$, regardless of the leads--QD coupling asymmetry parameter $\alpha$. 

In the following, we will consider the effects of temperature on the unhybridized system's linear conductance. In the finite temperature regime, the evaluation of the system's conductance requires numerical estimations based on the general Eqs.~\eqref{eq:12}-\eqref{eq:15}. Unlike the zero-temperature regime, in this case, all parameters in the problem are relevant to the evaluation.

Figure~\ref{fig:3} presents the system's total linear conductance as a function of temperature, $k_BT/\Gamma$, and the QD--MBS coupling strength, $|\lambda_2|/\Gamma$, for a symmetrical leads--QD coupling, $\alpha=1$, different values of the bias voltage asymmetry parameter ($q=0.25$ for Figs.~\ref{fig:3}a and \ref{fig:3}d, $q=0.5$ for Figs.~\ref{fig:3}b and \ref{fig:3}e, and $q=1$ for Figs.~\ref{fig:3}c and \ref{fig:3}f), and fixed second QD--MBS coupling strength ($|\lambda_1|/\Gamma=0.3$ for Figs.~\ref{fig:3}a--\ref{fig:3}c, and $|\lambda_1|/\Gamma=1.2$ for Figs.~\ref{fig:3}d--\ref{fig:3}f). For all figures, the MBSs are considered to be unhybridized, $\varepsilon_M=0$, the magnetic flux phase is $\phi=\pi$,  and the QD characteristic energy level is $\varepsilon_d=0$. In all considered cases, the temperature dependence of the total system's conductance is nonmonotonic. When the bias voltage asymmetry is $q=0.25$, as a function of temperature and QD--MBS coupling $|\lambda_2|/\Gamma$, the conductance presents a local maximum for the case of weak coupling strength, $|\lambda_1|/\Gamma=0.3$, which tends to occur at relatively low temperatures and low coupling strength $|\lambda_2|$. In the strong coupling strength case, $|\lambda_1|/\Gamma=1.2$, the maximum of the system's conductance seems to be less localized. This situation is characteristic of the system as long as the crossed Andreev reflection contribution to the total linear conductance is negative, i.e., $0<q<0.5$. When the crossed Andreev reflection switches to positive values, $0.5<q<1$, the local maximum still occurs at low temperatures, but in terms of its $|\lambda_2|$--dependence is less localized compared to the $0<q<0.5$ case.

\begin{figure*}[t]
	\includegraphics[width =1\linewidth]{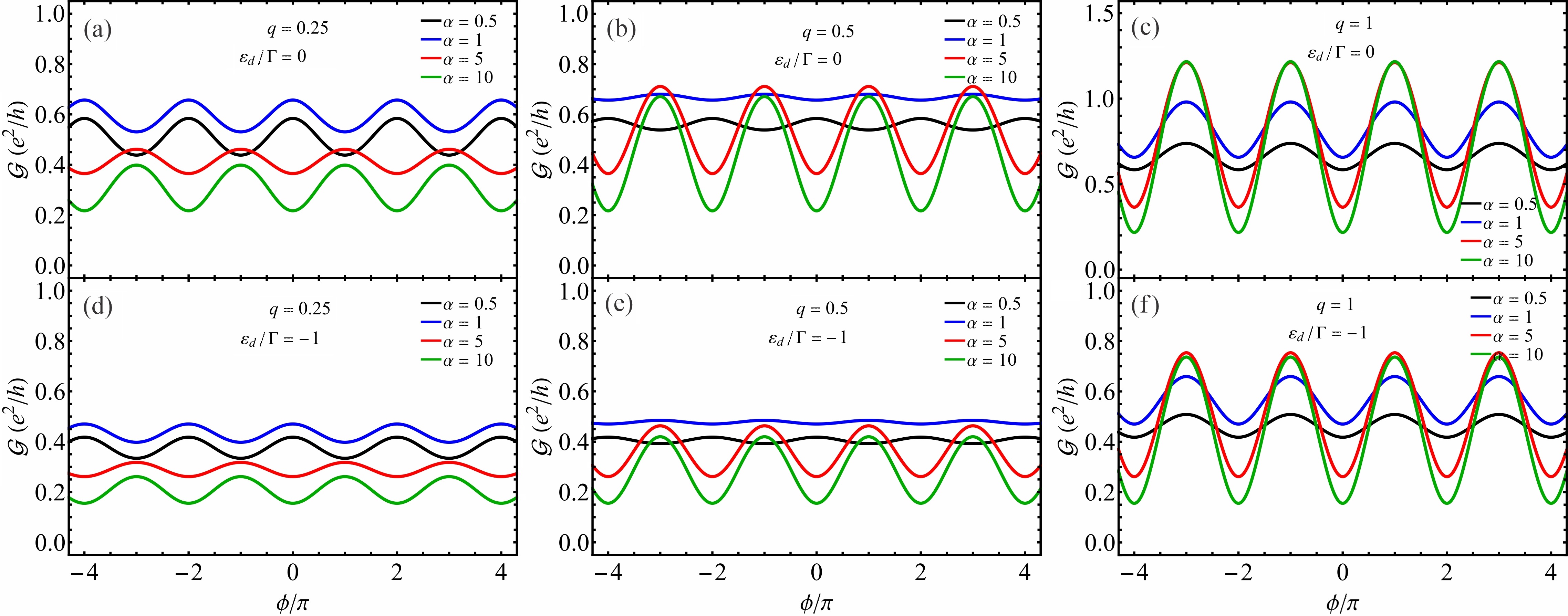}
	\centering
	\caption{The linear conductance $\mathcal{G}$ in the case of unhybridized MBSs ($\varepsilon_M/\Gamma=0$) as a function of magnetic flux phase $\phi$ for different values of the leads--QD coupling asymmetry parameter ($\alpha=0.5$ -- black line, $\alpha=1$ -- blue line, $\alpha=5$ -- red line, and $\alpha=10$ -- green line). The temperature is fixed at $k_B T/\Gamma = 0.3$ and the QD--MBS coupling strengths are $|\lambda_1|/\Gamma = |\lambda_2|/\Gamma = 0.3$. The QD characteristic energy is (a)--(c) $\varepsilon_d /\Gamma = 0$ and (d)--(f) $\varepsilon_d /\Gamma = -1$. The bias voltage asymmetry parameter is $q=0.25$ for (a) and (d),  $q=0.5$ for (b) and (e),  and  $q=1$ for (c) and (f).}
	\label{fig:6}
\end{figure*}

Figure~\ref{fig:4} presents the system's total linear conductance as a function of the leads--QD coupling asymmetry parameter $\alpha$ for different values of the bias voltage asymmetry parameter ($q=0.25$ for Figs.~\ref{fig:4}a and~\ref{fig:4}d, $q=0.5$ for Figs.~\ref{fig:4}b and~\ref{fig:4}e, and $q=1$ for Figs. \ref{fig:4}c and \ref{fig:4}f), and different temperature values ($k_BT/\Gamma=0$ -- black lines, $k_BT/\Gamma=0.3$ -- red lines, and $k_BT/\Gamma=1.2$ -- blue lines). For our calculations, we set the magnetic flux phase $\phi=\pi$, and the QD characteristic energy level $\varepsilon_d/\Gamma=0$ (solid lines) and $\varepsilon_d/\Gamma =-1$ (dashed lines). The MBSs are unhybridized ($\varepsilon_M=0$), and we consider two QD--MBSs coupling regimes, i.e., weak QD--MBSs coupling regime with $|\lambda_1|/\Gamma=0.3$ and $|\lambda_2|/\Gamma=0.1$ (Figs.~\ref{fig:4}a--\ref{fig:4}c), and strong QD--MBSs coupling regime with $|\lambda_1|/\Gamma=1.2$ and $|\lambda_2|/\Gamma=1$ (Figs.~\ref{fig:4}d--\ref{fig:4}f). As a general feature, in the finite-temperature regime, the system's linear conductance will depend on the QD characteristic energy and QD--MBSs coupling strengths. In the finite temperature and weak QD--MBSs coupling regime, when the asymmetry of the applied bias voltage is $q=0.25$ (Fig.~\ref{fig:4}a), for $\alpha<1$, the system's linear conductance increases, reaching a maximum value around $\alpha\simeq 1$, the limit of symmetric leads--QD coupling strengths. Further increase in $\alpha$ leads to a reduction in the system's linear conductance at finite temperature. This behavior is in stark contrast to the zero-temperature case, where the system's linear conductance increases for all values of $\alpha$; the higher the temperature, the larger the drop in conductance at large $\alpha$ values. In the low-temperature limit, the maximum system's conductance exceeds its value at $T=0$ for low and intermediate values of the leads--QD coupling asymmetry parameter $\alpha$ and weak QD--MBSs coupling strengths, $|\lambda_1|$ and $|\lambda_2|$. This general behavior is characteristic of the whole range $0<q<0.5$, when the contribution of the crossed Andreev reflection process to the system's linear conductance is negative. For $0.5<q<1$, the crossed Andreev reflection contribution to the system's linear conductance becomes positive. The general behavior of the system's total linear conductance is similar to that in the $0<q<0.5$ case, although in this case the conductance values exceed their corresponding values at $T=0$ only at low temperatures and in the low leads--QD coupling asymmetry parameter regime, $\alpha<1$, and weak QD--MBSs coupling strengths. In this case, for large values of the asymmetry parameter $\alpha$, the system's linear conductance still decreases as $\alpha$ increases, although the rate change is smaller than in the $0<q<0.5$ case. Interestingly, in the weak QD--MBSs coupling regime, for $\alpha=1$, the system's total linear conductance can exceed the zero-temperature Majorana fingerprint value $e^2/(2h)$ in the low-temperature limit, a result already reported for the symmetric bias voltage case $q=0.5$~\cite{Mathe2022}. In the strong QD--MBSs coupling regime, the linear conductance $\mathcal{G}$ increases monotonically with the leads--QD coupling asymmetry parameter $\alpha$ and tends to saturate at large $\alpha$ values. The temperature effect on the system's conductance is similar to that in the weak QD--MBSs coupling regime in the case of large leads--QD coupling asymmetry ($\alpha \gg 1$): as temperature increases, the system's total linear conductance decreases. For both the weak and strong QD--MBSs coupling regimes, considering a finite QD characteristic energy $\varepsilon_d/\Gamma =-1$ results in a reduction of the system's total linear conductance (dashed lines in Figs.~\ref{fig:4}a-\ref{fig:4}f correspond to $\varepsilon_d/\Gamma=-1$).  

\begin{figure*}[t]
	\includegraphics[width =0.95\linewidth]{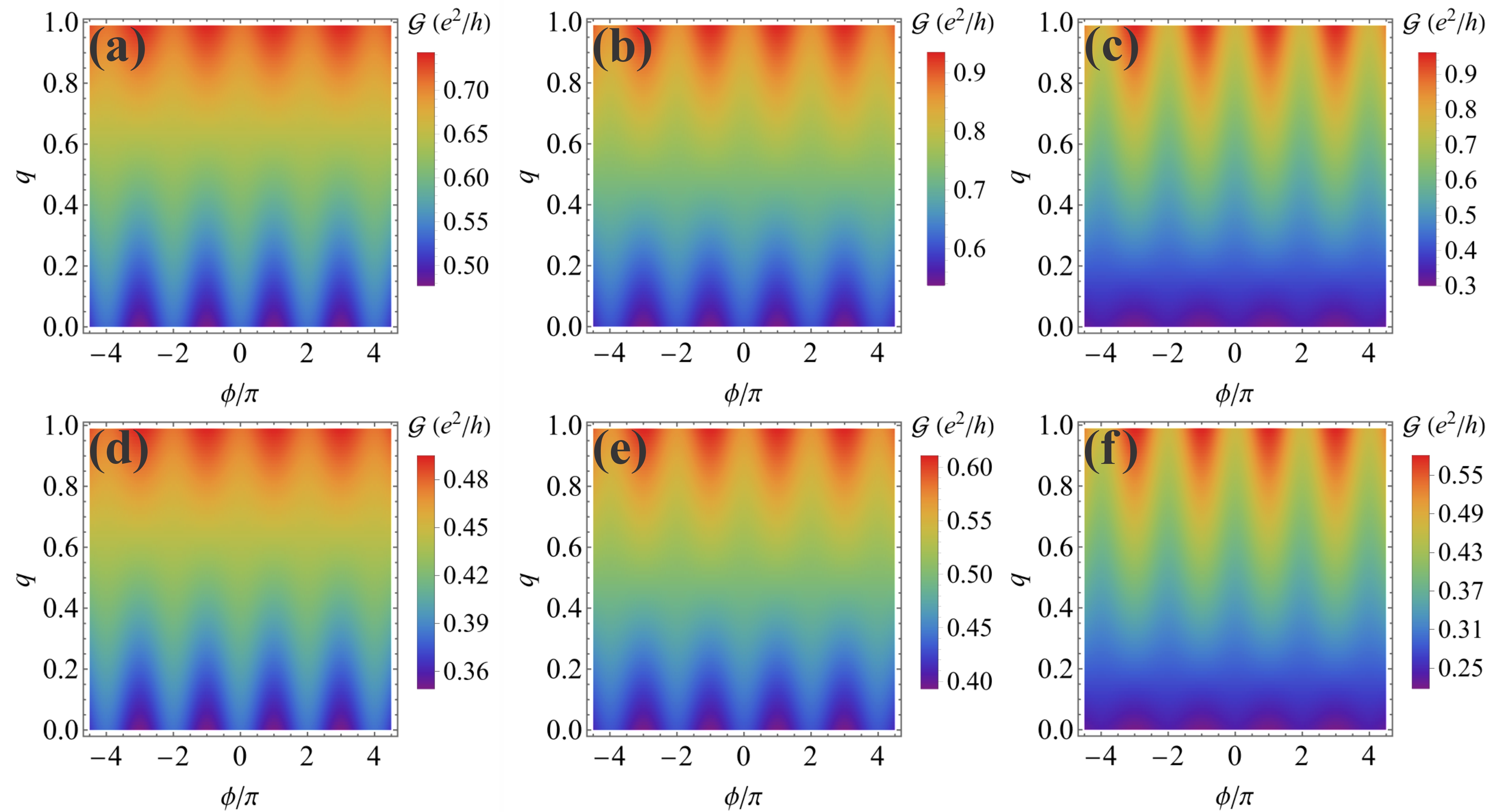}
	\centering
	\caption{The linear conductance $\mathcal{G}$ in the case of unhybridized MBSs ($\varepsilon_M/\Gamma=0$) as a function of magnetic flux phase $\phi$ and bias voltage asymmetry parameter $q$ for different values of the leads--QD coupling asymmetry parameter $\alpha$: (a) and (d) $\alpha=0.5$, (b) and (e) $\alpha=1$ and (c) and (f) $\alpha = 5$. The temperature is fixed at $k_B T/\Gamma = 0.3$ and the QD--MBSs coupling strengths are $|\lambda_1|/\Gamma = 0.3$ and $|\lambda_2|/\Gamma = 0.1$, respectively. The QD characteristic energy is (a)--(c) $\varepsilon_d /\Gamma = 0$ and (d)--(f) $\varepsilon_d /\Gamma = -1$.}
	\label{fig:7}
\end{figure*}

Figure~\ref{fig:5} presents the temperature dependence of the system's linear conductance for different values of the asymmetry parameter ($\alpha=0.5$ -- black line, $\alpha=1$ -- red line, and $\alpha=10$ -- blue line)  for the unhybridized MBSs case ($\varepsilon_M=0$) in the weak QD--MBSs coupling regime ($|\lambda_1|/\Gamma=0.3$ and $|\lambda_2|/\Gamma=0.1$). The magnetic flux phase is $\phi=\pi$. The asymmetry of the bias voltage is considered to be $q=0.25$ (Fig.~\ref{fig:5}a), $q=0.5$ (Fig.~\ref{fig:5}b), and $q=1$ (Fig.~\ref{fig:5}c), and the QD characteristic energy level is $\varepsilon_d=0$ (solid line) and $\varepsilon_d/\Gamma = -1$ (dashed line). As a function of temperature, the system's linear conductance presents a maximum in the low temperature limit when the leads--QD coupling asymmetry parameter $\alpha$ is in the low and intermediate range (below and in the vicinity of $\alpha \approx 1$). For larger values of $\alpha$, the maximum in the system's linear conductance shifts towards lower temperature values. In the high-temperature limit, the system's linear conductance decreases with temperature, independent of the bias voltage asymmetry parameter $q$. As a function of the QD characteristic energy level, one can see a decrease in the system's conductance when $\varepsilon_d$ is not tuned at the leads' Fermi level ($\varepsilon_d\neq 0$). For finite values of $\varepsilon_d$, the maximum in the system's conductance characteristic for the low and intermediate $\alpha$ regime moves towards larger temperatures and is diminished compared to the $\varepsilon_d=0$ case. The drop in the system's linear conductance when $\varepsilon_d\neq 0$ is characteristic of the whole range of the bias voltage asymmetry parameter $q$, and it tends to be more pronounced in the low and intermediate limit of the leads--QD coupling asymmetry parameter $\alpha$. In the case of a finite $\varepsilon_d$, the increase of the bias voltage asymmetry parameter $q$ tends to decrease the maximum linear conductance observed in the low temperature limit, even for the case of low or intermediate coupling asymmetries $\alpha$. For large leads--QD coupling asymmetry values $\alpha\rightarrow\infty$, the drop in the system's conductance related to finite values of $\varepsilon_d$ is reduced.

Figure~\ref{fig:6} presents the system's linear conductance in the case of unhybridized MBSs ($\varepsilon_M=0$) as a function of the magnetic flux phase $\phi$ for different values of the leads--QD coupling asymmetry parameter ($\alpha=0.5$ -- black line, $\alpha=1$ -- blue line, $\alpha=5$ -- red line, and $\alpha=10$ -- green line) in the finite temperature regime ($k_B T/\Gamma=0.3$). The QD characteristic energy level is $\varepsilon_d = 0$ (Figs.~\ref{fig:6}a--\ref{fig:6}c) and $\varepsilon_d/\Gamma=-1$ (Figs.~\ref{fig:6}d--\ref{fig:6}f), and the bias voltage asymmetry parameter $q = 0.25$ (Figs.~\ref{fig:6}a and \ref{fig:6}d), $q = 0.5$ (Figs.~\ref{fig:6}b and \ref{fig:6}e), and $q = 1$ (Figs.~\ref{fig:6}c and \ref{fig:6}f). The QD–MBSs coupling strengths are $|\lambda_1|/\Gamma = |\lambda_2|/\Gamma = 0.3$. The system's linear conductance presents a $2\pi$ periodicity as a function of the magnetic flux phase regardless of the value of the QD characteristic energy $\varepsilon_d$. As the value of the bias voltage asymmetry parameter $q$ increases, the amplitude of the conductance function increases, an effect that is stronger for larger values of the leads--QD coupling asymmetry parameter $\alpha$. A nonzero value of the QD characteristic energy $\varepsilon_d$  results in a reduction of the system's linear conductance.

Figure \ref{fig:7} presents the system's linear conductance in the case of unhybridized MBSs ($\varepsilon_M=0$) as a function of the magnetic flux phase $\phi$ and bias voltage asymmetry parameter $q$ for different values of the leads--QD coupling asymmetry parameter ($\alpha=0.5$ for Figs.~\ref{fig:7}a and~\ref{fig:7}d, $\alpha=1$ for Figs.~\ref{fig:7}b and~\ref{fig:7}e, and $\alpha=5$ for Figs.~\ref{fig:7}c and~\ref{fig:7}f). The central QD characteristic energy level $\varepsilon_d=0$ for Figs.~\ref{fig:7}a--\ref{fig:7}c and $\varepsilon_d/\Gamma=-1$ for Figs.~\ref{fig:7}d--\ref{fig:7}f. The QD--MBSs coupling strengths are $|\lambda_1|/\Gamma=0.3$ and $|\lambda_2|/\Gamma=0.1$, and the system's temperature is set at $k_B T/\Gamma=0.3$. As already mentioned, for unhybridized MBSs, the system's linear conductance presents a $2\pi$ periodicity as a function of the magnetic flux phase $\phi$, regardless of the value of the leads--QD asymmetry parameter $\alpha$ and characteristic energy level of the QD, $\varepsilon_d$. As a function of the magnetic flux phase $\phi$, the system's linear conductance maxima are located at $\phi=2n\pi$ ($n\in\mathbb{Z}$) when the bias voltage asymmetry parameter $q\rightarrow 0$. As the value of $q$ increases, the conductance maxima shift to $\phi=(2n+1)\pi$.  As a general feature, the maxima transition from $2n\pi$ to $(2n+1)\pi$ does not occur at a universal value of the bias voltage asymmetry parameter $q$,  but at $q$ values that will depend on the other properties of the system. Additionally, the larger the leads--QD coupling asymmetry parameter $\alpha$, the smaller the value of the bias voltage asymmetry parameter corresponding to the conductance maxima shift. This behavior results from the different magnetic-flux phase dependence of the system's conductance components, i.e., electron tunneling and the local and crossed Andreev reflection processes.

\begin{figure}[b]
	\includegraphics[width =0.98\linewidth]{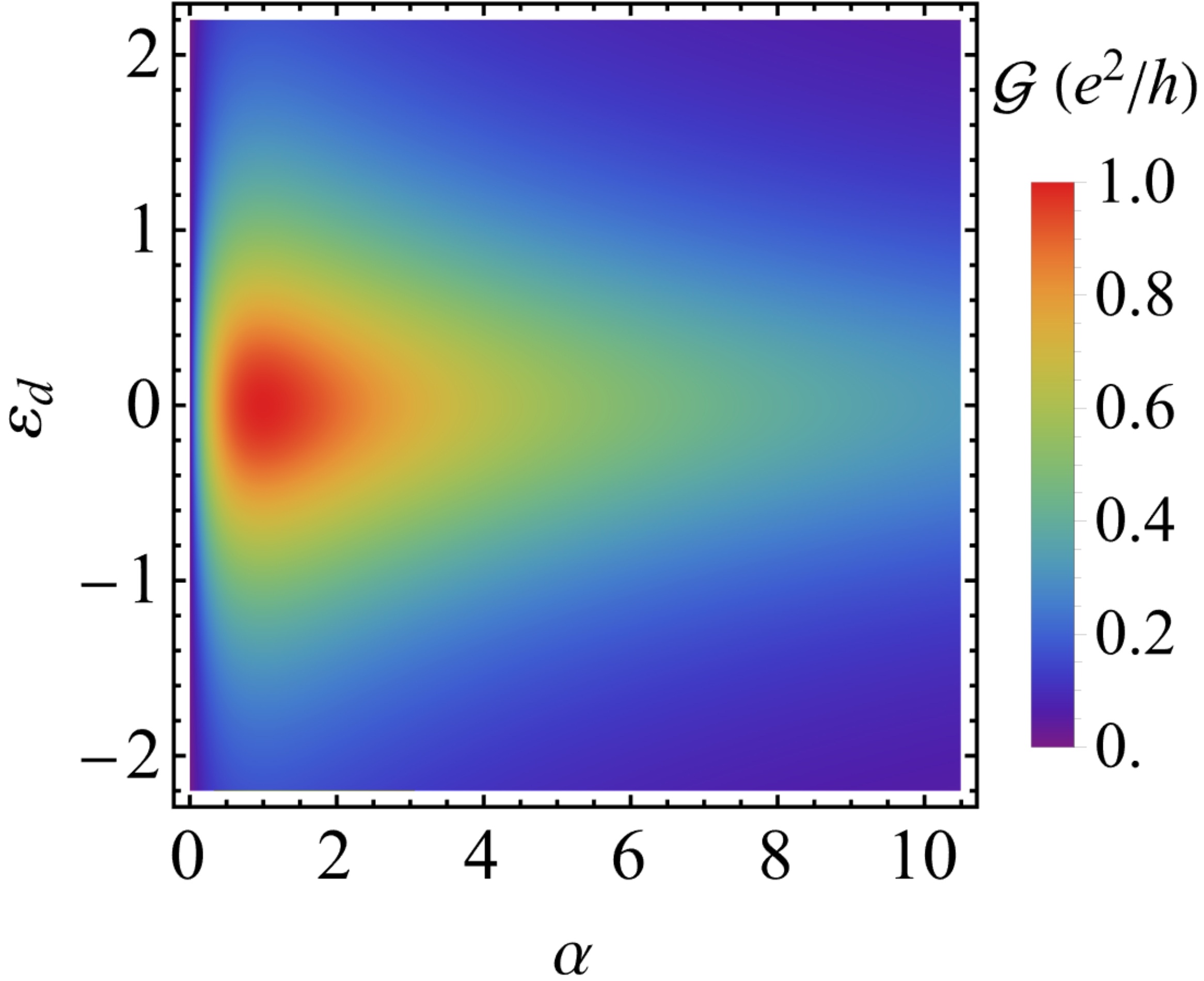}
	\centering
	\caption{The zero-temperature linear conductance $\mathcal{G}$ in the case of hybridized MBSs ($\varepsilon_M/\Gamma=0.3$) as a function of the leads--QD coupling asymmetry parameter $\alpha= \Gamma_L/\Gamma_R$ and the QD characteristic energy $\varepsilon_d$. The magnetic flux phase is $\phi=\pi$ and the QD--MBS coupling strengths are $|\lambda_1|/\Gamma=|\lambda_2|/\Gamma=0.3$.}
	\label{fig:8}
\end{figure}

\begin{figure*}[t]
	\includegraphics[width =0.85\linewidth]{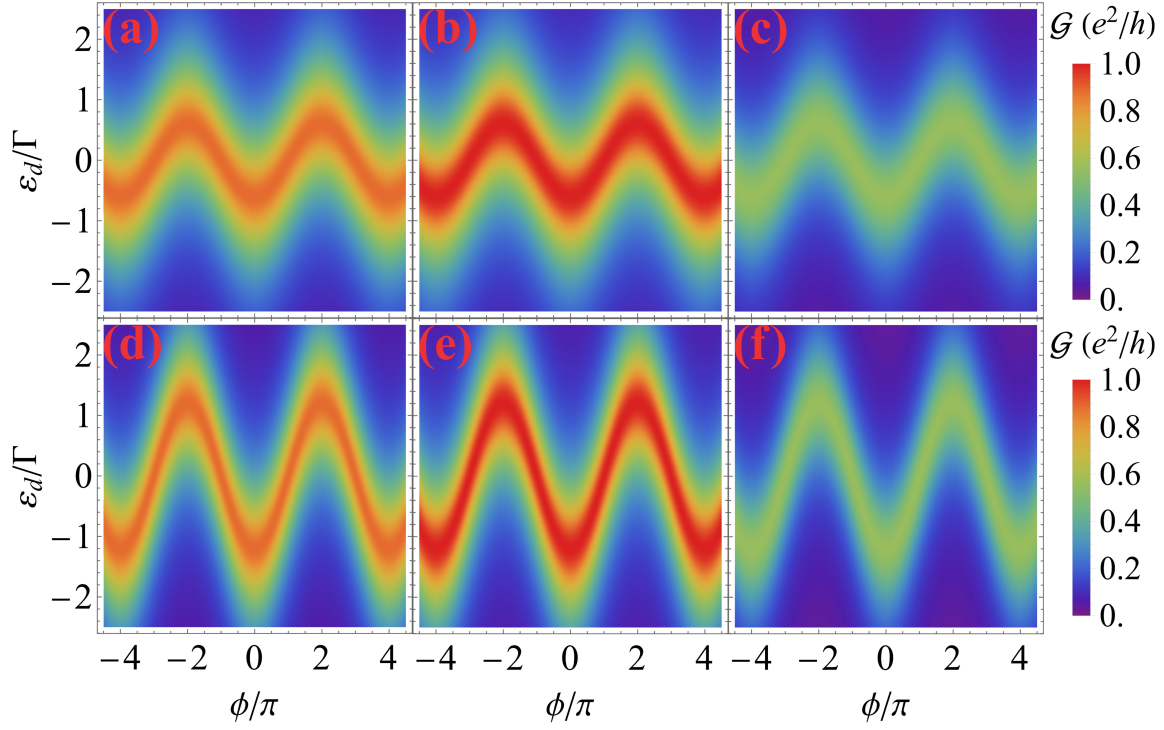}
	\centering
	\caption{The zero--temperature linear conductance $\mathcal{G}$ in the case of hybridized MBSs ($\varepsilon_M/\Gamma=0.3$) as a function of the magnetic flux phase $\phi$ and the QD characteristic energy $\varepsilon_d$. The QD--MBS coupling strengths are $|\lambda_1|/\Gamma = 0.3$, (a)-(c) $|\lambda_2|/\Gamma = 0.3$, and (d)-(f) $|\lambda_2|/\Gamma = 0.6$. The leads-QD coupling asymmetry parameter is $\alpha = 0.5$ for (a) and (d), $\alpha = 1$ for (b) and (e), and $\alpha = 5$ for (c) and (f).}
	\label{fig:9}
\end{figure*}

\subsection{Hybridized MBSs, \texorpdfstring{$\varepsilon_M\neq 0$}{EMF}}

Consider now the case of hybridized MBSs characterized by a finite Majorana overlap energy $\varepsilon_M\neq 0$. In the zero--temperature regime, the system's linear conductance, along with its components, depends on the values of the retarded Green's functions $G^r_{d11}(\varepsilon)$ and $G^r_{d12}(\varepsilon)$ in the $\varepsilon\rightarrow 0$ limit. A relatively simple estimation shows that in the case of hybridized MBSs, $G^r_{d12}(\varepsilon\rightarrow 0)=0$, and implicitly, the local and crossed Andreev reflection processes do not contribute to the total linear conductance of the system, implying that the electron tunneling component dominates the system's linear conductance, $\mathcal{G}=\mathcal{G}^{\text{ET}}$. As contributions from the Andreev reflection processes vanish in the zero--temperature regime, the system's conductance is independent of $q$, the asymmetry parameter associated with the applied bias voltage. Based on Eq.~\eqref{eq:13}, for finite values of $\varepsilon_M$ and for arbitrary values of $\phi$, the zero--temperature linear conductance of the system is given by
\begin{equation}
    \mathcal{G} = \frac{e^2}{h}\dfrac{4\,\alpha}{(\alpha + 1)^2}\frac{1}{\left(\dfrac{\varepsilon_d}{\Gamma}+\dfrac{2|\lambda_1 \lambda_2|}{\Gamma\,\varepsilon_M}\cos\dfrac{\phi}{2} \right)^2+1}.
    \label{eq:26}
\end{equation}
Clearly, in the hybridized MBSs case, even in the zero--temperature regime, the system's linear conductance depends on the ratio $\alpha=\Gamma_L/\Gamma_R$, the fixed total leads--QD coupling strength, $\Gamma$, the QD characteristic energy level, $\varepsilon_d$, as well as the MBSs characteristic parameters $\varepsilon_M$, $|\lambda_1|$, $|\lambda_2|$, and $\phi$. As a function of the leads--QD coupling asymmetry parameter $\alpha$, the system's linear conductance increases for $0<\alpha<1$, presents a maximum at $\alpha=1$, and eventually vanishes asymptotically for $\alpha\rightarrow\infty$. In stark contrast to the unhybridized case ($\varepsilon_M=0$), when the system's linear conductance asymptotically reaches a finite value as $\alpha\rightarrow\infty$, in the hybridized case ($\varepsilon_M\neq 0$), the system's linear conductance vanishes asymptotically in the same limit. This result is a consequence of the role played by the Andreev reflection processes in each case. Based on Eq. \eqref{eq:26}, in the zero--temperature regime, the system's conductance depends on the product of the two QD-MBSs coupling strengths, $|\lambda_1\lambda_2|$, and not separately on each individual QD-MBS coupling strength. Additionally, the system's total linear conductance is invariant under an $\alpha\rightarrow1/\alpha$ transformation.

\begin{figure*}[t]
	\includegraphics[width =1\linewidth]{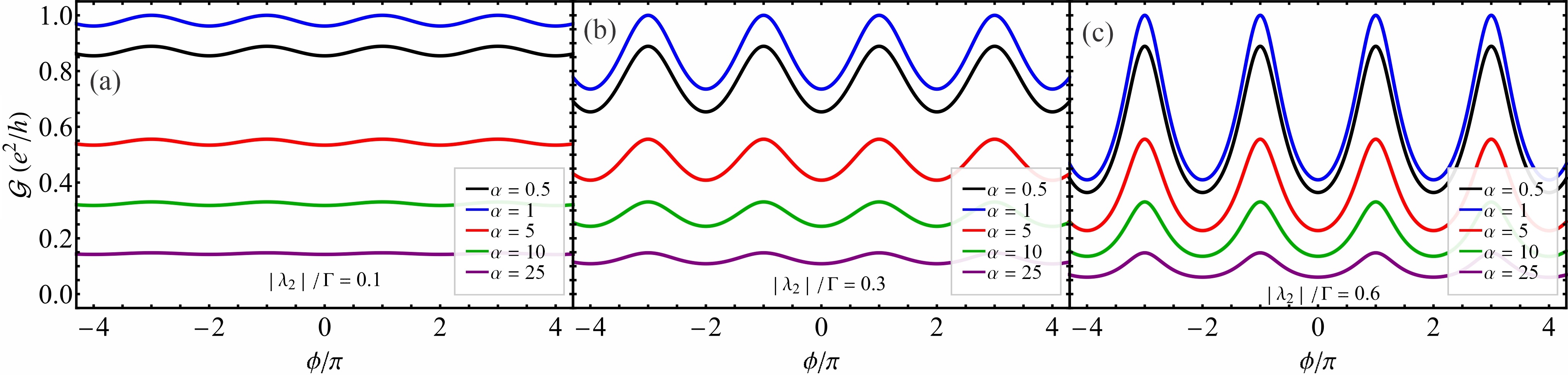}
	\centering
	\caption{The zero-temperature linear conductance $\mathcal{G}$ in the case of hybridized MBSs ($\varepsilon_M/\Gamma=0.3$) as a function of magnetic flux phase $\phi$ at different values of the leads--QD coupling asymmetry parameter ($\alpha=0.5$ -- black line, $\alpha=1$ -- blue line, $\alpha=5$ -- red line, $\alpha=10$ -- green line, and $\alpha=25$ -- purple line). The QD-MBSs coupling strengths are $|\lambda_1|/\Gamma = 0.3$ and (a) $|\lambda_2|/\Gamma = 0.1$, (b) $|\lambda_2|/\Gamma = 0.3$, and (c) $|\lambda_2|/\Gamma = 0.6$. The QD characteristic energy level is fixed at $\varepsilon_d = 0$.}
	\label{fig:10}
\end{figure*}

Figure~\ref{fig:8} presents the system's linear conductance as a function of the leads--QD coupling asymmetry parameter $\alpha$ and the QD characteristic energy $\varepsilon_d$. To focus on the role of the energy level in the QD, we consider the value of the magnetic flux phase to be $\phi=\pi$. In this case, the system's total linear conductance is an even function of $\varepsilon_d$ and presents a maximum at $\varepsilon_d=0$. For a fixed leads--QD coupling asymmetry parameter $\alpha$, the value of the total linear conductance diminishes as $|\varepsilon_d|$ increases. Figure~\ref{fig:9} presents the system's linear conductance as a function of the magnetic flux phase $\phi$ and QD characteristic energy $\varepsilon_d$ for different values of the leads--QD coupling asymmetry parameter $\alpha$ ($\alpha=0.5$ for Figs.~\ref{fig:9}a and~\ref{fig:9}d, $\alpha=1$ for Figs.~\ref{fig:9}b and~\ref{fig:9}e, and  $\alpha=5$ for Figs.~\ref{fig:9}c and \ref{fig:9}f). The MBSs overlapping energy is fixed at $\varepsilon_M/\Gamma=0.3$ and the QD-MBS coupling strength $|\lambda_1|/\Gamma=0.3$. The second QD-MBS coupling strength, $|\lambda_2|/\Gamma=0.3$ for Figs.~\ref{fig:9}a-\ref{fig:9}c, and $|\lambda_2|/\Gamma=0.6$ for Figs. \ref{fig:9}d-\ref{fig:9}f. As the magnetic flux phase value deviates from $\phi=(2n+1)\pi$, the maximum of the linear conductance shifts from the $\varepsilon_d=0$ value, and the even symmetry of the linear conductance function with respect to $\varepsilon_d$ is lost. Additionally, the $2\pi$ magnetic flux phase periodicity of the linear conductance at $\varepsilon_d=0$ shifts towards $4\pi$ as $\varepsilon_d\neq 0$~\cite{Mathe2022}. As expected, increasing the leads-QD coupling asymmetry parameter $\alpha$ modifies the amplitude and the width of the conductance resonances, while the position of the maxima remains unchanged. This indicates that the magnetic flux phase--induced interference condition is robust against variations in the tunneling coupling asymmetry.

Figure~\ref{fig:10} highlights the system's zero-temperature total linear conductance in the case of hybridized MBSs ($\varepsilon_M/\Gamma=0.3$) as a function of the magnetic flux phase $\phi$ for different leads--QD coupling asymmetry parameters ($\alpha=0.5$ -- black line, $\alpha=1$ -- blue line, $\alpha=5$ -- red line, $\alpha=10$ -- green line, and $\alpha=25$ -- purple line).  One of the  QD--MBS coupling strengths is fixed at $|\lambda_1|/\Gamma=0.3$, and the characteristic energy level in the central QD at $\varepsilon_d/\Gamma=0$. Figure~\ref{fig:10}a considers $|\lambda_2|<|\lambda_1|$ ($|\lambda_2|/\Gamma=0.1$), Figure~\ref{fig:10}b $|\lambda_2|=|\lambda_1|$ ($|\lambda_2|/\Gamma=0.3$), and Figure~\ref{fig:10}c $|\lambda_2|>|\lambda_1|$ ($|\lambda_2|/\Gamma=0.6$). As expected, in the case of hybridized MBSs, the linear conductance exhibits a $2\pi$ periodicity in the magnetic flux phase $\phi$, with maxima at $\phi=(2n+1)\pi$ ($n\in\mathbb{Z}$)~\cite{Mathe2022}. This periodicity is independent of the leads--QD coupling asymmetry parameter $\alpha$. However, as $\alpha$ increases, the linear conductance peaks become smeared, and the periodic dependence on the magnetic flux phase is no longer clearly resolved. There are a couple of effects on the system's total linear conductance related to the QD--MBSs coupling strengths $|\lambda_1|$ and $|\lambda_2|$. The minima values of the total linear conductance corresponding to the magnetic flux phase values $\phi = 2n\pi$, decrease significantly as the product $|\lambda_1\lambda_2|$ increases, while the maxima values at $\phi=(2n+1)\pi$ are not affected. This also implies sharper oscillations of the system's total linear conductance as a function of the magnetic flux phase $\phi$. This result suggests that one can compensate for the smeared conductance oscillations due to an increased leads--QD coupling asymmetry parameter $\alpha$ by increasing the QD--MBSs coupling strengths $|\lambda_1|$ and $|\lambda_2|$. Note that the system's linear conductance depends on the ratio $|\lambda_1\lambda_2|/\varepsilon_M$ (see eq.~\eqref{eq:26}), meaning that a similar effect can occur for fixed values of $|\lambda_1\lambda_2|$ and a variable $\varepsilon_M$. From the experimental point of view, this will result in more pronounced periodic modulations in the system's conductance with improved visibility of the periodic pattern. 

 \begin{figure*}[htb]
	\includegraphics[width =0.95\linewidth]{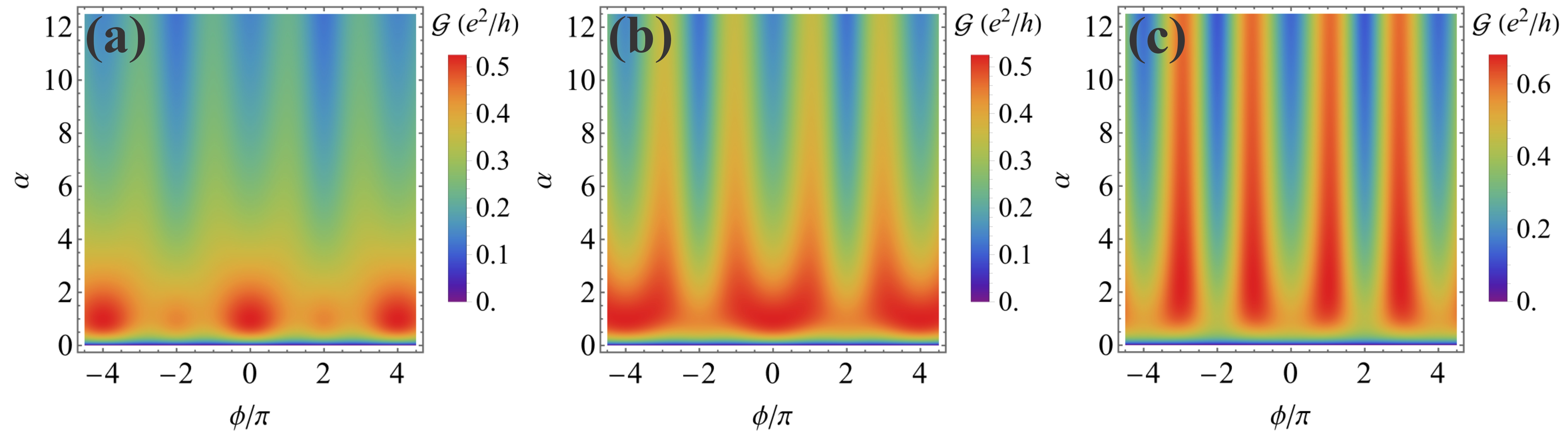}
	\centering
	\caption{The linear conductance $\mathcal{G}$ in the case of hybridized MBSs ($\varepsilon_M/\Gamma=0.3$) as a function of the magnetic flux phase $\phi$ and leads--QD coupling asymmetry parameter $\alpha$. The QD--MBSs coupling strengths are $|\lambda_1|/\Gamma=|\lambda_2|/\Gamma=0.3$, the QD characteristic energy is $\varepsilon_d/\Gamma = -1$, and the leads' temperature is set at $k_B T/\Gamma=0.3$. The bias voltage asymmetry parameter is (a) $q=0.25$, (b) $q=0.5$, and (c) $q=1$.}
	\label{fig:11}
\end{figure*}

Finally, let us consider the case in which one of the QD--MBSs coupling strengths vanishes, and the other is finite. Physically, this situation corresponds to the case when only one end of the Majorana nanowire is connected to the central QD. Based on Eq.~\eqref{eq:26}, the same result can be obtained when the magnetic flux phase is $\phi=(2n+1)\pi$, and both QD--MBSs coupling strengths are finite, regardless of the value of the leads--QD coupling strengths, $\alpha$, or bias voltage asymmetry parameters, $q$. This result suggests that tuning the magnetic flux phase threading the Majorana nanowire can provide insight into the transport properties of a simplified system comprising a single MBS coupled to the central QD.

In the following, we consider the temperature effects on the system's linear conductance for the case of hybridized MBSs ($\varepsilon_M\neq 0$). The evaluation of the system's linear conductance will be based on the general set of Eqs.~\eqref{eq:12}-\eqref{eq:15}. In this case, the full energy dependence of the two Green's functions $G^r_{d11}(\varepsilon)$ and $G^r_{d12}(\varepsilon)$ has to be considered, and consequently, different than in the zero--temperature case, both the local and crossed Andreev reflection processes contribute to the system's conductance. As a consequence, in the finite-temperature regime, the system's linear conductance will also depend on the bias voltage asymmetry parameter $q$.

Figure~\ref{fig:11} presents the system's linear conductance $\mathcal{G}$ in the case of hybridized MBSs ($\varepsilon_M/\Gamma=0.3$) as a function of magnetic flux phase $\phi$ and the leads--QD coupling asymmetry parameter $\alpha$ for different values of the bias voltage asymmetry parameter ($q=0.25$ -- Fig.~\ref{fig:11}a, $q=0.5$ -- Fig.~\ref{fig:11}b, and $q=1$ -- Fig.~\ref{fig:11}c). The QD--MBSs coupling strengths are considered in the weak coupling limit ($|\lambda_1|/\Gamma = |\lambda_2|/\Gamma = 0.3$), the QD characteristic energy is $\varepsilon_d/\Gamma=-1$, and the leads' temperature is set at $k_BT/\Gamma=0.3$. In the weak QD-MBSs coupling regime, as a function of the leads--QD coupling asymmetry parameter $\alpha$, the system's linear conductance presents a maximum at $\alpha=1$, and decreases as $\alpha$ increases for $\alpha>1$. As a function of the magnetic flux phase $\phi$, the system's linear conductance exhibits a $4\pi$ periodicity when $\varepsilon_d\neq 0$. In the finite-temperature limit, the system's linear conductance is also highly sensitive to the bias voltage asymmetry parameter $q$. In the small $q$ limit ($0<q<0.5$), the amplitude of the system's linear conductance is strongly diminished as the leads--QD coupling asymmetry $\alpha$ increases. In addition, due to the negative crossed Andreev reflection component of the linear conductance, as the value of the leads--QD coupling asymmetry parameter $\alpha$ diminishes towards $\alpha\rightarrow 0$,  the local maxima in the conduction function are shifted towards $\phi=4n\pi$. This effect is due to a destructive interference between the three components of the system's total linear conductance. In the large $q$ limit ($0.5<q<1$), when the magnetic flux is tuned at values corresponding to local maximum conductance points the amplitude of the linear conductance oscillations is enhanced, and the system's linear conductance seems to be less influenced by the leads--QD coupling asymmetry parameter $\alpha$.

\begin{figure*}[t]
	\includegraphics[width =1\linewidth]{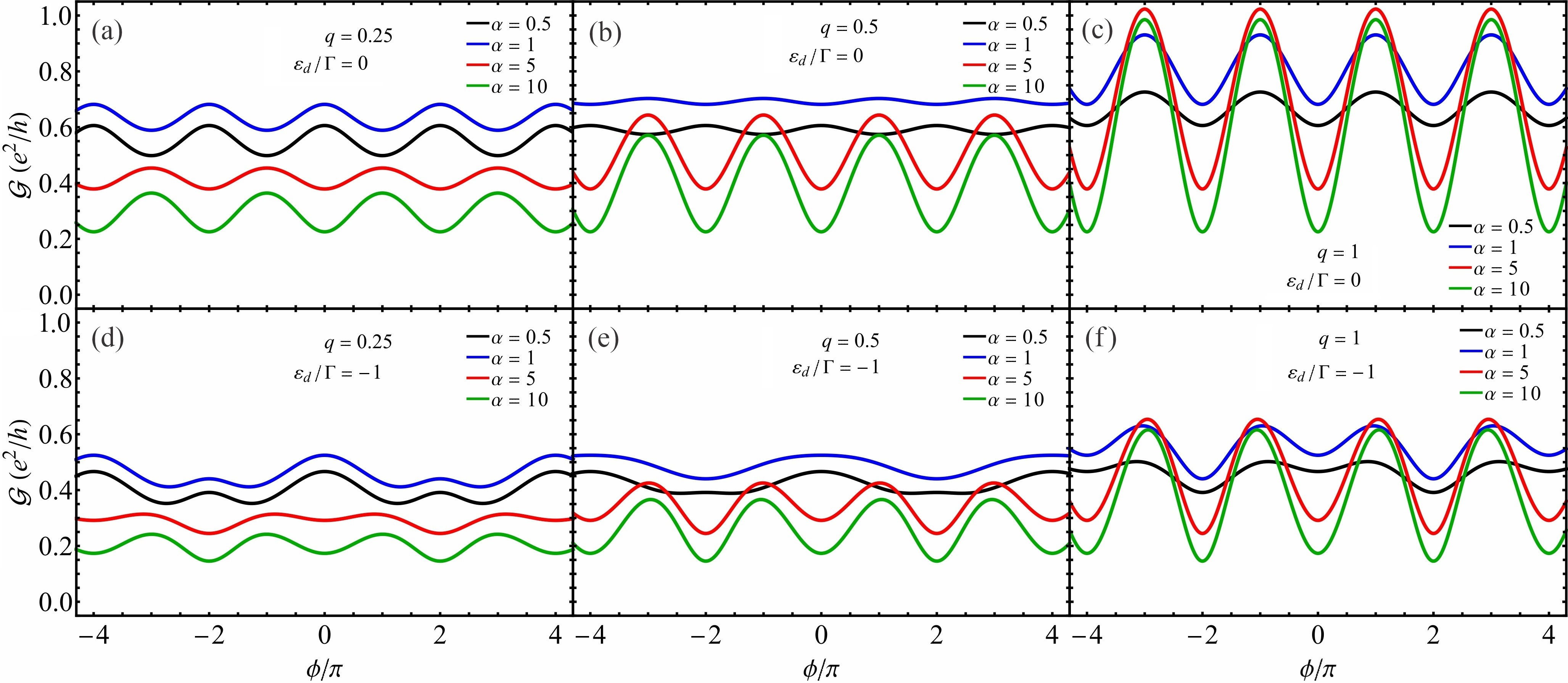}
	\centering
	\caption{The linear conductance $\mathcal{G}$ in the case of hybridized MBSs ($\varepsilon_M/\Gamma=0.3$) as a function of magnetic flux phase $\phi$ for different values of the leads--QD coupling asymmetry parameter ($\alpha=0.5$ -- black line, $\alpha=1$ -- blue line, $\alpha=5$ -- red line, and $\alpha=10$ -- green line).  The temperature is fixed at $k_B T/\Gamma = 0.3$ and the QD--MBSs coupling strengths are $|\lambda_1|/\Gamma = |\lambda_2|/\Gamma = 0.3$. The QD characteristic energy is (a)--(c) $\varepsilon_d /\Gamma = 0$ and (d)--(f) $\varepsilon_d /\Gamma = -1$. The bias voltage asymmetry parameter $q=0.25$ for (a) and (d),  $q=0.5$ for (b) and (e),  and  $q=1$ for (c) and (f).}
	\label{fig:12}
\end{figure*}

Figure~\ref{fig:12} presents the system's linear conductance as a function of the magnetic flux phase $\phi$ for different values of the leads--QD asymmetry parameter ($\alpha=0.5$ - black line, $\alpha=1$ - blue line, $\alpha=5$ - red line, and $\alpha=10$ - green line) in the finite temperature regime ($k_BT/\Gamma=0.3$). The QD characteristic energy level is $\varepsilon_d=0$ (Figs.~\ref{fig:12}a--\ref{fig:12}c) and $\varepsilon_d/\Gamma=-1$ (Figs.~\ref{fig:12}d--\ref{fig:12}f), and the bias voltage asymmetry parameter $q=0.25$ (Figs.~\ref{fig:12}a and~\ref{fig:12}d), $q=0.5$ (Figs.~\ref{fig:12}b and \ref{fig:12}e), and $q=1$ (Figs. \ref{fig:12}c and~\ref{fig:12}f). The QD--MBSs coupling strengths are $|\lambda_1|/\Gamma=|\lambda_2|/\Gamma=0.3$. In stark contrast to the zero-temperature limit (see Fig.~\ref{fig:10}b), the system's linear conductance at finite temperature depends on the bias voltage asymmetry parameter $q$. As the value of the bias voltage asymmetry parameter increases, the amplitude of the linear conductance oscillations is strongly enhanced, in particular for higher values of the leads--QD coupling asymmetry parameter $\alpha$. Similar to the zero--temperature regime, the system's linear conductance oscillation period is $2\pi$ for $\varepsilon_d=0$ and changes to $4\pi$ when $\varepsilon_d\neq 0$ (see Figs.~\ref{fig:12}d--\ref{fig:12}f for $\varepsilon_d/\Gamma=-1$).

\begin{figure*}[t]
	\includegraphics[width =0.95\linewidth]{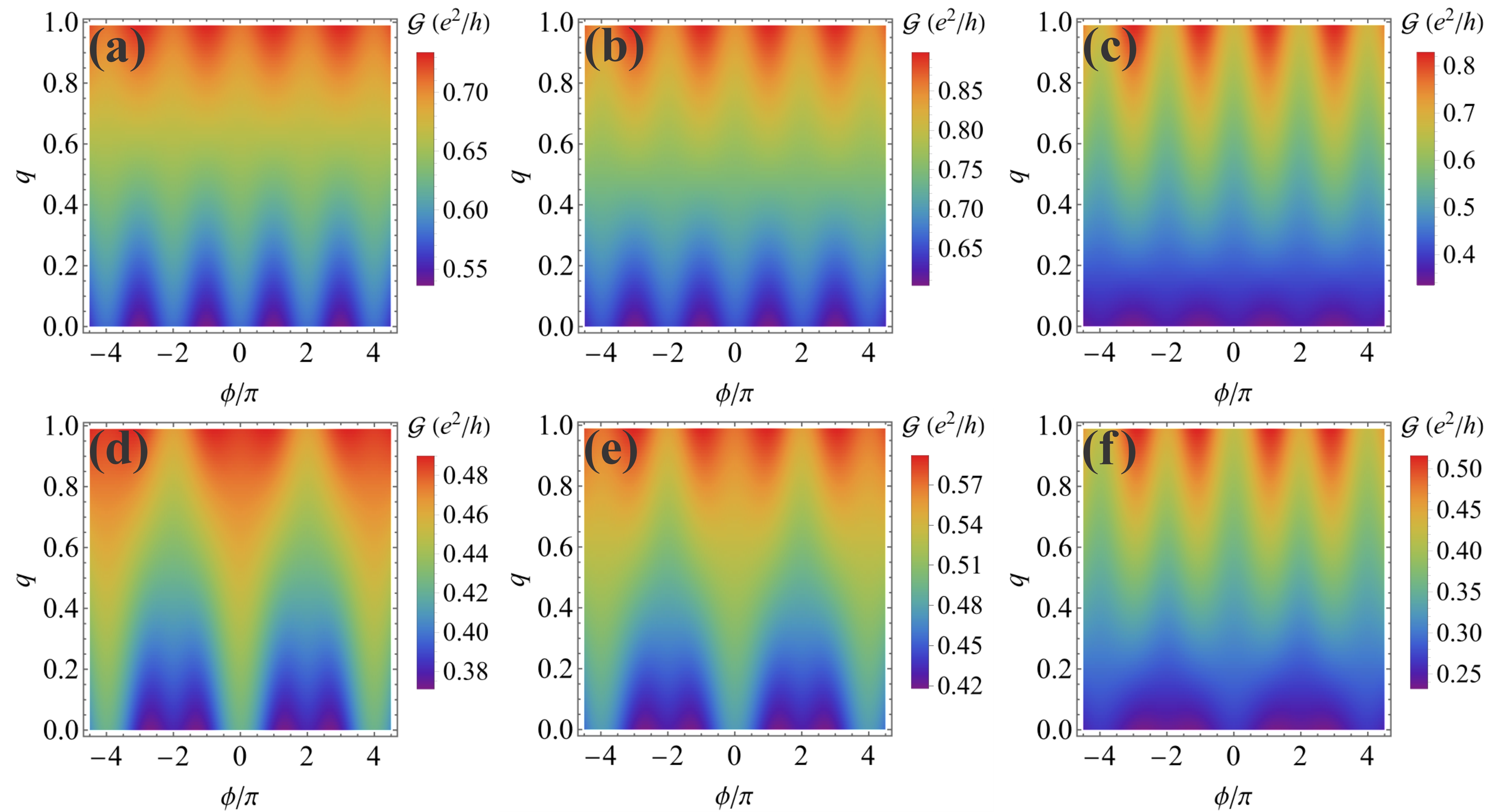}
	\centering
	\caption{The linear conductance $\mathcal{G}$ in the case of hybridized MBSs ($\varepsilon_M/\Gamma=0.3$) as a function of magnetic flux phase $\phi$ and bias voltage asymmetry parameter $q$ for different values of the leads--QD coupling asymmetry parameter $\alpha$: (a) and (d) $\alpha=0.5$, (b) and (e) $\alpha=1$ and (c) and (f) $\alpha = 5$. The temperature is fixed at $k_B T/\Gamma = 0.3$ and the QD--MBS coupling strengths are $|\lambda_1|/\Gamma = 0.3$ and $|\lambda_2|/\Gamma = 0.1$, respectively. The QD characteristic energy is (a)--(c) $\varepsilon_d /\Gamma = 0$ and (d)--(f) $\varepsilon_d /\Gamma = -1$. }
	\label{fig:13}
\end{figure*}

Figure~\ref{fig:13} highlights the system's linear conductance in the case of hybridized MBSs ($\varepsilon_M/\Gamma=0.3$) as a function of the magnetic flux phase $\phi$ and bias voltage asymmetry parameter $q$ for different values of the leads--QD coupling asymmetry parameter ($\alpha=0.5$ for Figs.~\ref{fig:13}a and ~\ref{fig:13}d, $\alpha=1$ for Figs.~\ref{fig:13}b and ~\ref{fig:13}e, and  $\alpha = 5$ for Figs.~\ref{fig:13}c and ~\ref{fig:13}f). The central QD characteristic energy level $\varepsilon_d=0$ for Figs. ~\ref{fig:13}a--~\ref{fig:13}c and $\varepsilon_d/\Gamma=-1$ for Figs.~\ref{fig:13}d--~\ref{fig:13}f. The QD--MBSs coupling strengths are $|\lambda_1|/\Gamma = 0.3$ and $|\lambda_2|/\Gamma = 0.1$, and the system's temperature is set at $k_B T/\Gamma = 0.3$. In the hybridized case, the periodicity of the system's linear conductance as a function of the magnetic flux phase $\phi$ is $2\pi$ when the central QD characteristic energy is $\varepsilon_d=0$ and $4\pi$ when the central QD characteristic energy is $\varepsilon_d\neq 0$. When $\varepsilon_d=0$,  the linear conductance maxima are located at $2n\pi$ for low values of the bias voltage asymmetry parameter $q$ and switch to $(2n+1)\pi$ when the bias voltage asymmetry $q$ increases. When $\varepsilon_d\neq 0$, in the low bias voltage asymmetry limit, the linear conductance maxima are located at $4n\pi$ regardless of the value of the leads--QD coupling asymmetry parameter $\alpha$. However, as the bias voltage asymmetry parameter $q$ increases, one can notice a shift in the local linear conductance maxima points, although their exact localization as a function of the magnetic flux phase $\phi$ will strongly depend on the other parameters in the system. This characteristic of the system's linear conductance is more prominent for low and intermediate values of the leads--QD coupling asymmetry parameter $\alpha$, although it is also visible for large values of the leads--QD coupling asymmetry parameter.

\begin{figure*}[t]
	\includegraphics[width =0.85\linewidth]{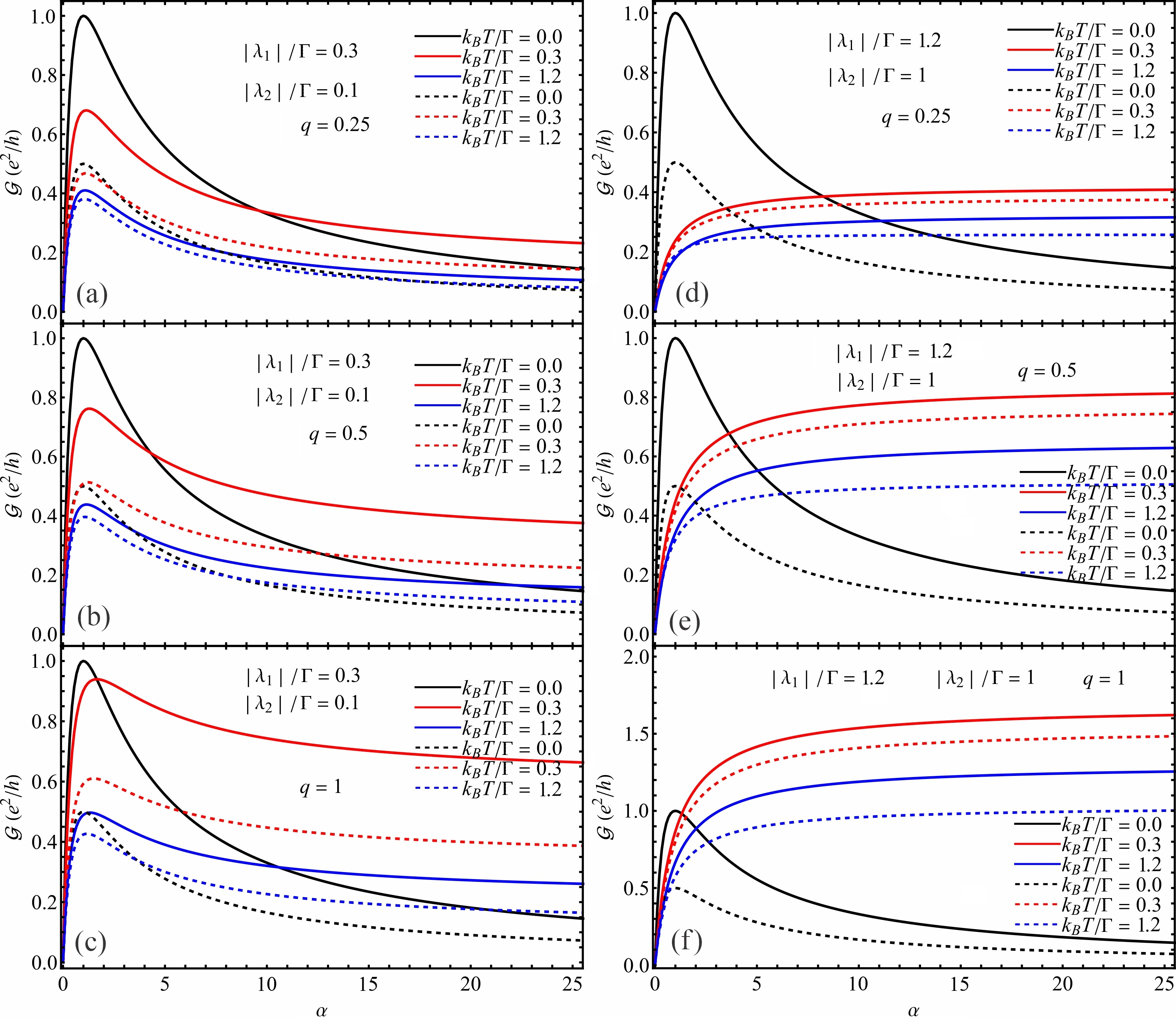}
	\centering
	\caption{The linear conductance $\mathcal{G}$ in the case of hybridized MBSs ($\varepsilon_M/\Gamma=0.3$) as a function of the leads--QD coupling asymmetry parameter $\alpha$ for different values of the temperature ($k_BT/\Gamma=0$ -- black line, $k_BT/\Gamma=0.3$ -- red line, and $k_BT/\Gamma=1.2$ -- blue line). The QD--MBSs coupling strengths are (a)--(c) $|\lambda_1|/\Gamma = 0.3$ and $|\lambda_2|/\Gamma=0.1$, and (d)--(f) $|\lambda_1|/\Gamma = 1.2$ and $|\lambda_2|/\Gamma=1$. The bias voltage asymmetry parameter is $q=0.25$ for (a) and (d), $q=0.5$ for (b) and (e),  and $q=1$ for (c) and (f). The magnetic flux phase is $\phi=\pi$ and the central QD characteristic energy $\varepsilon_d/\Gamma = 0$ (solid lines) and $\varepsilon_d/\Gamma = -1$ (dashed lines).}
	\label{fig:14}
\end{figure*}

Figure~\ref{fig:14} presents the system's total linear conductance as a function of the leads--QD coupling asymmetry parameter $\alpha$ for different values of the bias voltage asymmetry parameter ($q=0.25$ for Figs.~\ref{fig:14}a and~\ref{fig:14}d, $q=0.5$ for Figs.~\ref{fig:14}b and~\ref{fig:14}e, and $q=1$ for Figs. \ref{fig:14}c and \ref{fig:14}f), and different values of the temperature ($k_BT/\Gamma=0$ - black lines, $k_BT/\Gamma=0.3$ - red lines, and $k_BT/\Gamma=1.2$ - blue lines). For our calculations, we set the magnetic flux phase $\phi=\pi$, and we considered two different values for the QD characteristic energy level $\varepsilon_d/\Gamma=0$ (solid lines) and $\varepsilon_d/\Gamma=-1$ (dashed lines). The MBSs are hybridized ($\varepsilon_M/\Gamma=0.3$), and we consider the weak QD--MBSs coupling regime with $|\lambda_1|/\Gamma=0.3$ and $|\lambda_2|/\Gamma=0.1$ (Figs.~\ref{fig:14}a-\ref{fig:14}c), and in the strong QD-MBSs coupling regime with $|\lambda_1|/\Gamma=1.2$ and $|\lambda_2|/\Gamma=1$ (Figs.~\ref{fig:14}d-\ref{fig:14}f). The main temperature effects are highlighted in Fig. \ref{fig:14}; in the large $\alpha$ limit ($\alpha\gg 1$), compared to the zero--temperature case, one can see an increase in the system's linear conductance due to additional contributions from the Andreev reflection processes. A larger conductance value is consistent in the strong QD--MBSs coupling regime for all bias voltage asymmetries $q$. On the other hand, in the weak QD--MBSs coupling regime, the system's linear conductance decreases at large values of the coupling asymmetry parameter $\alpha\gg 1$, and the temperature enhancement of the overall conductance is smaller. In particular, the temperature effects are noticeably smaller for bias voltage asymmetries $0<q<0.5$ when the crossed Andreev reflection contribution to the system's conductance is negative.

Similar to the unhybridized case, in the weak QD-MBSs coupling regime, for all values of the bias voltage asymmetry parameter $q$ (Fig.~\ref{fig:14}), when $\alpha<1$, the system's linear conductance increases as a function of $\alpha$ at finite temperature, reaching a maximum value around $\alpha\simeq 1$, the limit of symmetrically coupled QD. Further increase in $\alpha$ leads to a reduction in the system's linear conductance at finite temperature. For finite temperatures, in the strong QD--MBSs coupling regime, the linear conductance $\mathcal{G}$ increases monotonically with the leads--QD coupling asymmetry parameter $\alpha$  and tends to saturate at large $\alpha$ values. Due to the additional contribution of the Andreev reflection processes, the system's linear conductance is always larger than the zero--temperature conductance in the large $\alpha$ limit. The effect of the QD characteristic energy $\varepsilon_d\neq 0$ for both the weak and strong QD-MBSs coupling regimes is a reduction of the system's linear conductance compared to its value when $\varepsilon_d=0$ (dashed lines in Figs.~\ref{fig:14}a-\ref{fig:14}f correspond to $\varepsilon_d/\Gamma=-1$). 

\begin{figure*}[t]
	\includegraphics[width =1\linewidth]{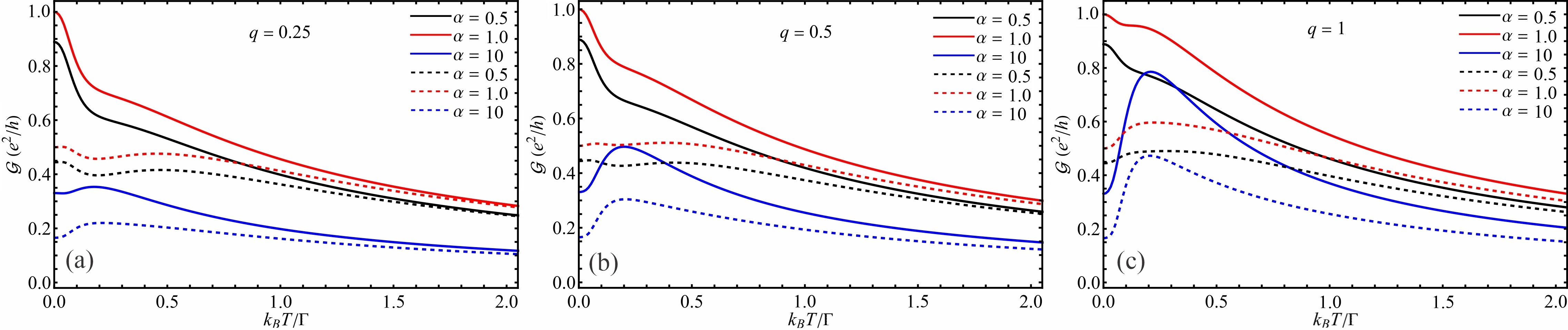}
	\centering
	\caption{The total linear conductance $\mathcal{G}$ in the case of hybridized MBSs ($\varepsilon_M/\Gamma =0.3$) as a function of temperature $k_B T/\Gamma$ at different values of the leads--QD coupling asymmetry parameter ($\alpha=0.5$ -- black line, $\alpha=1$ -- red line, and $\alpha=10$ -- blue line) and different values of the bias voltage asymmetry parameter (a) $q=0.25$, (b) $q=0.5$ and (c) $q=1$. The QD-MBSs coupling strengths are $|\lambda_1|/\Gamma= 0.3$ and $|\lambda_2|/\Gamma = 0.1$, and the magnetic flux phase is $\phi=\pi$. The solid and dashed lines correspond to the QD characteristic energy level $\varepsilon_d/\Gamma = 0$ and $\varepsilon_d/\Gamma = -1$, respectively.}
	\label{fig:15}
\end{figure*}

Figure~\ref{fig:15} presents the temperature dependence of the system's linear conductance in the case of hybridized MBSs ($\varepsilon_M/\Gamma=0.3$) as function of temperature for different values of the leads--QD coupling asymmetry parameter ($\alpha=0.5$ - black line, $\alpha=1$ - red line, and $\alpha=10$ - blue line) in the weak QD--MBSs coupling regime ($|\lambda_1|/\Gamma=0.3$ and $|\lambda_2|/\Gamma=0.1$). The magnetic flux phase is $\phi=\pi$, the bias voltage asymmetry parameter is $q=0.25$ (Fig.~\ref{fig:15}a), $q=0.5$ (Fig.~\ref{fig:15}b), and $q=1$ (Fig.~\ref{fig:15}c), and the QD characteristic energy level is $\varepsilon_d=0$ (solid line) and $\varepsilon_d/\Gamma=-1$ (dashed line). In the low--temperature limit, the system's linear conductance exhibits different features depending on the values of the asymmetry parameters $q$ and $\alpha$. Most notably, for bias voltage asymmetry parameters $0.5<q<1$, the finite temperature conductance can exceed its zero--temperature value even for large values of the leads--QD coupling asymmetry parameter $\alpha$. For large values of $\alpha$, a local maximum in the linear conductance function can occur at low temperatures (see Fig.~\ref{fig:15}c for $q=1$ and $\alpha=10$). As a general feature, in the hybridized case, the system's linear conductance decreases with increasing temperature in the high-temperature limit. When the QD characteristic energy $\varepsilon_d$ is tuned away from the leads' Fermi energy, $\varepsilon_d\neq 0$, the system's linear conductance decreases. This change is visible even in the zero--temperature regime: in the hybridized case, the system's linear conductance depends on $\varepsilon_d$ at $T=0$. The dashed lines in Fig. \ref{fig:15} highlight this effect for $\varepsilon_d/\Gamma=-1$.

\section{Summary}
\label{sec:IV}

In this work, we studied the charge quantum transport through a QD system that is coupled to both external leads and two MBSs arising from a topological superconducting nanowire. The topological superconducting nanowire and the central QD form a closed ring that is threaded by a tunable magnetic flux. The system's transport properties are governed by two physical processes: electrons can tunnel between the external leads under an applied bias voltage, and they can undergo local and crossed Andreev reflection due to the presence of the superconducting nanowire. We focused on the quantum transport when the leads--QD coupling strengths and applied bias voltage are asymmetric, and we considered two different scenarios, when the MBSs are unhybridized or hybridized, in the zero and finite temperature regimes.

In the case of unhybridized MBSs ($\varepsilon_M=0$) for the zero--temperature regime, all three processes, the electron tunneling, the local Andreev reflection, and the crossed Andreev reflection, contribute to the system's linear conductance. As a general result, the system's linear conductance vanishes as long as the magnetic flux phase in the Majorana nanowire is not an odd multiple of $\pi$ ($\phi\neq (2n+1)\pi$, $n\in\mathbb{Z}$). Interestingly, in this case, the system's linear conductance does not depend on the value of the central QD characteristic energy $\varepsilon_d$, and the QD-MBSs coupling strengths $|\lambda_1|$ and $|\lambda_2|$. As a function of the leads--QD coupling asymmetry parameter $\alpha=\Gamma_L/\Gamma_R$ and the applied bias voltage asymmetry parameter $q$, the system's linear conductance scales as $\mathcal{G}/\mathcal{G}_0=2q\alpha/(\alpha+1)$ ($\mathcal{G}_0=e^2/h$), a result that generalized the Majorana signatures already discussed for the symmetric coupling case ($\alpha=1$ and $q=0.5$). For large values $\alpha\rightarrow\infty$, the system's linear conductance is dominated by the local Andreev reflection process. In the finite temperature regime, in the case of unhybridized MBSs, the system's linear conductance also depends on the central QD characteristic energy $\varepsilon_d$, and the QD--MBSs coupling strengths $|\lambda_1|$ and $|\lambda_2|$. In particular, as a function of the leads--QD coupling asymmetry parameter $\alpha$, we identified a nonmonotonic behavior of the system's linear conductance in the limit of weak QD--MBSs coupling. Quite differently, in the strong QD--MBSs coupling limit, the system's linear conductance increases monotonically as $\alpha$ increases both at zero and finite temperature. When the central QD characteristic energy is tuned away from the leads' Fermi energy level, $\varepsilon_d\neq 0$, the system's linear conductance is reduced, although it maintains a similar behavior as in the $\varepsilon_d=0$ situation.

The situation differs for hybridized MBSs ($\varepsilon_M\neq 0$). Our analysis assumes $\varepsilon_M/\Gamma=0.3$, although varying this ratio yields similar results \cite{Mathe2022}. In the zero-temperature regime, the system's linear conductance is unaffected by local or crossed Andreev reflection processes and arises solely from electron tunneling between the left and right leads. Accordingly, the quantum transport in the system is not affected by the bias voltage asymmetry parameter $q$, and vanishes in the large leads--QD coupling asymmetry parameter limit $\alpha\rightarrow\infty$. In stark contrast to the unhybridized MBSs case, even in the zero--temperature regime, the system's linear conductance is largely influenced by the central QD characteristic energy $\varepsilon_d$, QD--MBSs coupling strengths $|\lambda_1|$ and $|\lambda_2|$, and the magnetic flux phase $\phi$. When the central QD characteristic energy is $\varepsilon_d=0$, as a function of the magnetic phase flux, the system's linear conductance presents a $2\pi$ periodicity, with conductance peaks smearing out as the leads--QD coupling asymmetry parameter increases. However, our results suggest that one can compensate for the smeared conductance oscillations amplitudes by increasing the QD--MBSs coupling strengths $|\lambda_1|$ and $|\lambda_2|$. On the other hand, when the central QD characteristic energy is tuned away from the leads' Fermi energy level, $\varepsilon_d\neq 0$, the periodicity of the linear conductance oscillations changes to $4\pi$. In the finite temperature regime, all three processes, the electron tunneling, the local Andreev reflection, and the crossed Andreev reflection, contribute to the system's linear conductance for all nonzero values of the bias voltage asymmetry parameter $q\neq 0.5$. Consequently, in this regime, the bias voltage asymmetry parameter $q$ plays an important role in the system's transport properties. In particular, as $q$ increases, the system's linear conductance presents sharper peaks as a function of the magnetic flux phase $\phi$ for all values of the leads--QD coupling asymmetry parameter $\alpha$. On the other hand, as a function of the leads--QD coupling asymmetry parameter $\alpha$, the system's linear conductance presents a nonmonotonic behavior in the weak QD--MBSs coupling regime, with conductance values larger than their corresponding values in the zero--temperature regime when $\alpha \gg 1$. This behavior changes to a monotonic one in the strong QD--MBSs coupling regime. Similar to the zero--temperature regime, as the central QD characteristic energy level is shifted away from the leads' Fermi level, the linear conductance decreases, and its oscillations shift from $2\pi$ to $4\pi$. In general, higher temperatures correspond to lower conductance values for both the unhybridized and hybridized MBSs cases.
  
A very interesting situation arises in the finite-temperature regime for both the unhybridized and hybridized Majorana cases. As we already mentioned, the system's linear conductance has three contributions. The weight of these contributions to the total linear conductance depends on the bias voltage and the coupling asymmetry parameters, $q$ and $\alpha$. The electron tunneling contribution is independent of $q$ and depends on $\alpha$, $\mathcal{G}^{\text{ET}}\propto \alpha/(\alpha+1)^2$, the local and crossed Andreev reflection contributions they both depend on $q$ and $\alpha$, $\mathcal{G}^{\text{LAR}}\propto 2q\alpha^2/(\alpha+1)^2$ and $\mathcal{G}^{\text{CAR}}\propto (2q-1)\alpha/(\alpha+1)^2$. As a function of the magnetic flux phase $\phi$, the local and crossed Andreev reflection contributions will have a similar behavior, as both depend on $|G^r_{d12}(\varepsilon)|^2$. The behavior might differ for the electron tunneling contribution, as the magnetic flux phase dependence of this component is related to $|G^r_{d11}(\varepsilon)|^2$. In the case of unhybridized MBSs ($\varepsilon_M=0$), independent of the value of the central QD characteristic energy level $\varepsilon_d$ and the value of the leads--QD coupling asymmetry parameter $\alpha$, the system's linear conductance presents a $2\pi$ periodicity as a function of the magnetic flux phase $\phi$. As a function of the magnetic flux phase $\phi$, the linear conductance maxima occur at $2n\pi$ ($n\in\mathbb{Z}$) as long as the bias voltage asymmetry parameter $q\rightarrow 0$, and shift towards $\phi=(2n+1)\pi$ as the value of $q$ increases. The position of the linear conductance maxima changes at nonuniversal $q$ values that depend on other properties of the system as a result of the interference between the three components of the system's linear conductance. As a general feature, the bias voltage transition point decreases as the value of the leads--QD coupling asymmetry parameter $\alpha$ increases. This situation is similar for hybridized MBSs ($\varepsilon_M\neq 0$), as long as the central QD characteristic energy is tuned at the Fermi level ($\varepsilon_d=0$).  For the case of hybridized MBSs, when the central QD characteristic energy is tuned away from the leads' Fermi levels, $\varepsilon_d\neq 0$, the conductance periodicity as a function of the magnetic flux phase is $4\pi$.  In this case, for low bias voltage asymmetry $q$, the maxima of the linear conductance occur at $4n\pi$ ($n\in\mathbb{Z}$), and as $q$ increases, they shift to a different $\phi$ location, although this location is not universal and it strongly depends on the system's properties.

The results presented in this work should be relevant to current efforts to realize and control MBSs in nanoscale QD systems~\cite{DasSarma2023,Kouwenhoven2025}. Recent advances in nanoscale QD systems fabrication provide the experimental basis for the practical realization of QD-based topological superconducting nanowire junctions~\cite{Deng2016,Deng2018,Lutchyn2018,Gao2024,Shabani2016,Whiticar2020,Heedt2021}. In particular, controlling such a system via asymmetric leads-QD coupling strengths or applied bias voltages, or via easily accessible and tunable magnetic fluxes, can be an important step toward developing scalable and topologically protected Majorana qubit architectures~\cite{Flensberg2011,Alicea2012,Hyart2013,Landau2016,Qin2019,Razmadze2020}.

\begin{acknowledgments}
The authors would like to thank Dr.~Doru~Sticleț for valuable discussions. L. M. and L. P. Z. gratefully acknowledge financial support through the ``Nucleu'' Program within the National Research Development and Innovation Plan 2022–2027, Romania, carried out with the support of MEC, project no. 27N/03.01.2023, component project code PN 23 24 01 04, and from CNCS/CCCDI-UEFISCDI, under project number PN-IV-P1-PCE-2023-0987.
\end{acknowledgments}

\bibliographystyle{apsrev4-2}
\bibliography{References}
\end{document}